\documentstyle[12pt,psfig]{article}
\newif
\iffigs
\figstrue     

\def\ms1{{\kern .035em s}}
\def\tnl{t_{\rm nl}}
\def\ds{\displaystyle}
\makeatletter \def\fps@figure{tp} \makeatother
\iffigs
\fi
\def\drawing #1 #2 #3 {
\begin{center}
\setlength{\unitlength}{1mm}
\begin{picture}(#1,#2)(0,0)
\put(0,0){\framebox(#1,#2){#3}}
\end{picture}
\end{center} }

\begin{document}
\title{On the decay of Burgers turbulence}
\author{S.N. Gurbatov$^{1,2}$, S.I. Simdyankin$^{1,3}$, E. Aurell$^{3,4}$,
U. Frisch$^2$ \& G. T\'oth$^{5,6}$}
\date{}
\maketitle
\begin{tabular}{ll}
$^1$ &Radiophysics Dept., University of Nizhny Novgorod,\\
& 23, Gagarin Ave., Nizhny Novgorod 603600, Russia. \\
&(Permanent address) \\ \\
$^2$ &Observatoire de la C\^ote d'Azur, Lab. G.D. Cassini,\\
& B.P. 4229, F-06304 Nice Cedex 4, France.\\\\
$^3$ & Center for Parallel Computers,\\
& Royal Institute of Technology,\\
& S--100 44 Stockholm, Sweden.\\\\
$^4$ & Mathematics Dept., Stockholm University,\\
& S--106 91 Stockholm, Sweden.\\\\
$^5$ &Dept. Atomic Physics, Lor\'and E\"otv\"os University,\\
&Budapest 1088, Puskin u. 5--7, Hungary.\\\\
$^6$ &Sterrekundig Instituut,\\
& Postbus 80000, 3508 TA Utrecht, The Netherlands.\\\\
\end{tabular}

\centerline{{\it J. Fluid Mech. 1997, \bf 344}, 339--374}
\centerline{(Received 9 December 1996 and in revised form 16 April 1997)}
\newpage
 \begin{abstract} \noindent This work is devoted to the decay of
random solutions of the unforced Burgers equation in one dimension in
the limit of vanishing viscosity. The initial velocity is homogeneous
and Gaussian with a spectrum proportional to $k^n$ at small
wavenumbers $k$ and falling off quickly at large wavenumbers.

In physical space, at sufficiently large distances, there is an ``outer
region'', where the velocity correlation function preserves exactly its initial
form (a power law) when $n$ is not an even integer.

When $1<n<2$ the spectrum, at long times, has three scaling regions\,:
first, a $|k|^n$ region at very small $k$\ms1 with a time-independent
constant, stemming from this outer region, in which the initial
conditions are essentially frozen; second, a $k^2$ region at
intermediate wavenumbers, related to a self-similarly evolving ``inner
region'' in physical space and, finally, the usual $k^{-2}$ region,
associated to the shocks. The switching from the $|k|^n$ to the $k^2$
region occurs around a wavenumber $k_s(t) \propto t^{-1/[2(2-n)]}$,
while the switching from $k^2$ to $k^{-2}$ occurs around
$k_L(t)\propto t^{-1/2}$ (ignoring logarithmic corrections in both
instances).

When $-1<n<1$ there is no inner region and the long-time evolution of 
the spectrum is self-similar. 

When $n=2,4,6,\ldots$ the outer region disappears altogether and the
long-time evolution is again self-similar. For other values of $n>2$,
the outer region gives only subdominant contributions to the small-$k$
spectrum and the leading-order long-time evolution is also
self-similar.

The key element in the derivation of the results is an extension of
the Kida (1979) log-corrected $1/t$ law for the energy decay when
$n=2$ to the case of arbitrary integer or non-integer $n>1$. A
systematic derivation is given in which both the leading term and
estimates of higher order corrections can be obtained. It makes use of
the Balian--Schaeffer (1989) formula for the distribution in space of
correlated particles, which gives the probability of having a given
domain free of particles in terms of a cumulant expansion. The
particles are, here, the intersections of the initial velocity
potential with suitable parabolas. The leading term is the Poisson
approximation used in previous work, which ignores correlations.

High-resolution numerical simulations are presented which support our
findings.
 \end{abstract}
\newpage
\section{Introduction}
\label{s:intro}

The Burgers equation
\begin{equation}
\partial_t v  +v\partial_x v= \nu \partial_x^2 v,
\label{burgers}
\end{equation}
describes a variety of nonlinear wave phenomena arising in the theory
of wave propagation, acoustics, plasma physics and so on (see, e.g.,
Whitham 1974; Gurbatov, Malakhov \& Saichev 1991). It was originally
introduced by Jan~M.~Burgers (1939) as a toy model for turbulence.  On
the one hand, it shares a number of properties with the Navier--Stokes
equation\,: the same type of nonlinearity, of invariance groups and of
energy-dissipation relation, the existence of a multidimensional version,
etc.  On the other hand, it is known to be integrable (Hopf 1950; Cole
1951) and therefore lacks the property of sensitive dependence on the
initial conditions (chaos).  When this was realized, interest in the
Burgers equation as a toy model for Navier--Stokes turbulence abated
somewhat. More recently, it became clear that a number of questions
connected to the Burgers equation are far from trivial. One class of
problems, arising e.g. in surface growth (Barab\'asi \& Stanley 1995),
leads to a Burgers equation with a random driving force (usually
white-noise in time) added to the right hand side (Kardar, Parisi \&
Zhang 1986; Bouchaud, M\'ezard \& Parisi 1995; Cheklov \& Yakhot 1995;
Polyakov 1995; Gurarie \& Migdal 1996; E {\it et al.} 1997).  Even in
the absence of a driving force, the explicit solution provided by the
Hopf--Cole method leads to non-trivial problems in the limit of
vanishing viscosity, when random initial conditions are assumed. An
example, motivated by work on large-scale structures in the Universe
(Zeldovich 1970; Gurbatov \& Saichev 1984; Gurbatov, Saichev \&
Shandarin 1989; Shandarin \& Zeldovich 1989), is when the initial
velocity field is a Brownian (or fractional Brownian) process in the
space variable (She, Aurell \& Frisch 1992; Sinai 1992; Vergassola
{\it et al.} 1994; Aurell {\it et al.} 1997). Other examples arise in
the sudy of disordered systems (Bouchaud \& M\'ezard 1997).

Another instance, which is at the core of the present
study, is the long-time behavior of  freely decaying solutions in the
limit of zero viscosity and, in particular the law of
decay of the  energy. Before reviewing this issue it is useful to put
it into a broader  perspective, namely the law of decay for
three-dimensional Navier--Stokes turbulence.

K\'arm\'an \& Howarth (1938; see also Lin \& Reid 1963), investigating
the decay of high-Reynolds number, homogeneous isotropic
three-dimensional turbulence, proposed a self-preservation
(self-similarity) ansatz for the spatial correlation function of the
velocity\,: the correlation function keeps a fixed functional shape;
the integral scale $L(t)$ and the root-mean-square velocity $u(t)$
have a long-time power-law behavior; there are two exponents which can
be related by the condition that the energy dissipation per unit mass
should be proportional to $u^3/L$.  But {\em an additional relation\/}
is needed to actually determine the exponents. Kolmogorov (1941)
realized this and proposed using the so-called ``Loitsyansky (1939)
invariant'' as the additional relation and, thereby, derived a law of
energy decay $u^2(t) \propto t^{-10/7}$. Proudman \& Reid (1954) and
Batchelor \& Proudman (1956) showed that the integral considered by
Loitsyansky is actually not invariant.  Under the assumption that the
energy spectrum $E(k,t)\propto k^4$ at small wavenumbers $k$, the
Loitsyansky integral is, within a numerical factor, just the
coefficient of $k^4$.  For isotropic turbulence, the contribution of
the nonlinear term to the time derivative of the energy spectrum,
which is called the transfer function, should be proportional to $k^4$
at small $k$, thereby preventing the constancy of the Loitsyansky
integral. However, if initially $E(k)\propto |k|^n$ with a ``spectral
exponent'' $n<4$, then the coefficient of $|k|^n$ will be
invariant. This is called here the principle of ``permanence of large
eddies'' (PLE). Kolmogorov's argument is easily adapted to that case
and gives then $u^2(t) \propto t^{-2(n+1)/(n+3)}$ (see, e.g., Frisch
1995, Section~7.7).

Almost exactly the same arguments may be applied to the
one-dimensional Burgers turbulence. In one dimension the transfer
function should be proportional to $k^2$ at small $k$ (see
Section~\ref{s:pheno}). Hence, if the initial energy spectrum
$E_0(k)\propto |k|^n$ with $-1 < n < 2$, Kolmogorov's argument predicts
again $u^2(t) \propto t^{-2(n+1)/(n+3)}$.  We shall see that this is
correct only when $-1 < n < 1$.

For Navier--Stokes turbulence, a rigorous derivation of the law of
energy decay seems ruled out at present.\footnote{It is not even
clear, in three dimensions, if for the initial value problem there is
any energy decay at all in the limit of vanishing viscosity, since the
issue of breakdown of smoothness for the Euler flow is moot (see,
e.g., Frisch 1995, Section~7.8).} The situation is of course more
favorable for Burgers-type turbulence and it will be our main goal in
this paper to critically examine the issues of self-similarity and
energy decay at long times. In this paper we shall exclusively
consider the behavior of the solutions in the limit of vanishing
viscosity for Gaussian initial data. For the effect of a finite (small)
viscosity and/or non-Gaussian initial data (e.g., piecewise constant or
shot noise velocities), the reader is referred to recent papers by
Funaki, Surgailis \& Woyczynski (1995), Surgailis (1996, 1997) and Hu \&
Woyczynski (1996).

The paper is organized as follows. In Section~\ref{s:formulation}, we
formulate our problem and list some elementary results about the Burgers
equation to be used in what follows.  Section~\ref{s:pheno} is devoted
to phenomenological derivations of the main results about the decay of
random solutions to the Burgers equation, the nature of which depends
crucially on the spectral exponent $n$. The most delicate results,
obtained when $n>1$, are rederived by systematic methods in
Sections~\ref{s:homogeneous} and~\ref{s:correlations}.  In
Section~\ref{s:previouswork} we discuss previous work on this
issue. In Section~\ref{s:homogeneous} we derive the law of decay of
energy as the leading term in a systematic expansion. Our basic tool
is the cumulant expansion (Balian \& Schaeffer, 1989) of the
probability of finding regions in space without particles.  In
Section~\ref{s:correlations} we calculate the spatial correlation
function, and prove the permanence of large eddies. In
Section~\ref{s:intermediate} we summarize our main findings and derive
consequences, such as the absence of globally self-similar decay for
the energy spectrum when $1<n<2$.  Section~\ref{s:numerics} presents
high-resolution numerical experiments which confirm such findings.
Section~\ref{s:conclusion} presents concluding remarks.
The Appendix contains a rederivation of the Balian--Schaeffer
formula in a form appropriate for our problem.

\section{Formulation and elementary results}
\label{s:formulation}

We shall be concerned with the initial-value problem for the unforced
Burgers equation in one dimension\,:
\begin{equation}
\partial_t v  +v\partial_x v= \nu \partial_x^2 v, \qquad v_0(x)\equiv v(x,0).
\label{BE}
\end{equation}
Use will also be made of the conservation form of the Burgers equation\,:
\begin{equation}
\partial_t v =  \partial_x
\left( -\frac{v^2}{2} + \nu \partial_x v \right)
\label{BEC}
\end{equation}
and of the formulation in terms of the (velocity) potential\,:
\begin{equation}
\partial_t \psi = \frac{1}{2} (\partial_x \psi)^2 +
\nu \partial_{xx} \psi,
\label{BEP}
\end{equation}
where
\begin{equation}
v = - \partial_x \psi.
\label{vpsi}
\end{equation}
As is well-known, the solution to the Burgers equation with positive
viscosity $\nu$, has an explicit integral representation obtained by
Hopf (1950) and Cole (1951).  We are mainly interested here in
the solutions in the limit of vanishing viscosity. Use of Laplace's
method then leads to the following ``maximum representation'' for the
potential (Hopf 1950) in the limit of vanishing viscosity (henceforth
always understood)\,:
\begin{eqnarray}
\psi (x,t) &=& \max_a \Phi(x,a,t)\label{MAX}\\
\Phi(x,a,t) &\equiv& \left[ \psi_0(a) - \frac{(x-a)^2 } {2t} \right],
\label{Phifunction}
\end{eqnarray}
where $\psi_0(a)$ is the initial potential. Expansion of the square
shows that (\ref{MAX}) is basically a Legendre transformation. From
(\ref{vpsi}) and (\ref{MAX}) it follows that
\begin{equation}
v (x,t) = \frac{ x-a(x,t) } {t} ,
\label{velocity111}
\end{equation}
where $ a(x,t) $ is the coordinate at which $\Phi(x,a,t)$ achieves its
(global) maximum  for given $x$ and $t$. It is easily checked that $a$
is the Lagrangian coordinate from which emanates the fluid particle
which will be at $x$ at time $t$ (see, e.g., Gurbatov {\it et al.} 1991).

In what follows we shall work with homogeneous random velocity fields.
We list the main  definitions. The correlation function is
\begin{equation}
B_v(x,t)\equiv \langle v(x,t) v(0,t)\rangle,
\label{defbv}
\end{equation}
where the angular brackets denote ensemble averages.
Its Fourier transform, the energy spectrum is
\begin{equation}
E(k,t) \equiv \frac{1}{2\pi} \int_{-\infty}^{\infty}
B_v(x,t) \exp(ikx)\, dx.
\label{defek}
\end{equation}
The initial values of the correlation function and of the energy
spectrum  are denoted $B_{0v}(x,t)$ and $E_0(k)$, respectively.
The (mean) energy is
\begin{equation}
E(t)\equiv \langle v^2(x,t)\rangle = \int_{-\infty}^\infty E(k,t)\,dk.
\label{defet}
\end{equation}
The initial energy is denoted
\begin{equation}
\sigma_v^2 \equiv \langle v_0^2(x)\rangle = \int_{-\infty}^\infty E_0(k)\,dk.
\label{defsigmav}
\end{equation}

We now introduce the basic assumptions of this paper. The initial
velocity field is homogeneous and Gaussian with an energy spectrum of
the form
\begin{equation}
E_0(k) = \alpha^2 |k|^n b_0(k),
\label{initspecpowerlaw}
\end{equation}
where $n$ is the spectral exponent. The even and non-negative function
$b_0(k)$ is assumed to satisfy $b_0(0) = 1$. It is smooth and
decreases faster than any inverse power of $k$ at infinity (this
implies, of course, that the initial velocity correlation function is
very smooth).  $b_0(k)$
can be characterized by a wavenumber $k_0$ around which lies most of
the initial energy and which is, in order of magnitude, the inverse of
the initial integral scale $L_0$.  The usual definition of the
integral scale, which involves the integral of the correlation
function, is not appropriate since this integral vanishes for any
$n>0$.  Instead, we use
\begin{equation}
L_0 \equiv {\sigma_\psi\over\sigma_v},
\label{defL0}
\end{equation}
where $\sigma_{\psi}^2$ is the variance of the initial potential
\begin{equation}
\sigma_{\psi}^2 \equiv \langle \psi_0^2(x)\rangle =
\int_{-\infty}^\infty {E_0(k)\over k^2}\,dk.
\label{defsigmapsi}
\end{equation}
This definition of  $L_0$ will be convenient in deriving various long-time laws in 
Sections~\ref{s:homogeneous} and \ref{s:correlations}.

As we shall see, the value of the spectral exponent $n$ determines the
universality class of the decaying solution at long times. In order
for the velocity to be homogeneous its variance must be finite; hence,
we must have $n>-1$. When $-3<n<-1$ the velocity has infinite variance
and only increments are homogeneous; the velocity itself behaves, at
large distances, as a Brownian or fractional Brownian motion process,
a case considered by She {\it et al.} (1992), Sinai (1992),
Vergassola {\it et al.} (1994), which is not within the scope of the
present paper. When $-1<n<1$ the velocity is homogeneous whereas the
potential behaves, at large distances, as a Brownian or fractional
Brownian motion process. This case has already been investigated by
Burgers (1974), by She {\it et al.} (1992) and by Avellaneda \& E
(1995) and will be considered here only briefly. We shall concentrate
on the case $n>1$, when both the velocity and the potential are
homogeneous.

\subsection{Elementary results}
\label{s:elementary}

We observe that the smoothness of the energy spectrum at the origin
depends on the spectral exponent $n$. When $n$ is an even integer, our
assumption implies that the initial correlation function $B_{0v}(x)$,
being the $n$th derivative of a smooth function of $x$ decreasing
rapidly at infinity, also has these properties. When $n$ is odd or
non-integer the spectrum is not smooth. It follows that the correlation
function decreases as a power law (Pitman 1968; Bleistein \& Handelsman 1975),
{\it viz}\,\footnote{In this paper the symbol $\simeq$ is used for the leading-order term in
asymptotic expansions, $\sim$ is used to relate two
quantities which may differ by an order unity dimensionless constant
and $\propto$ when the quantities are just proportional.}
\begin{equation}
B_{0v}(x) \simeq \alpha^2C_n |x|^{-n-1}, \qquad |x|\to\infty,
\label{largexinit}
\end{equation}
where 
\begin{equation}
C_n= -2\Gamma(n+1)\, \sin {\pi n\over2}.
\label{Cnvalue}
\end{equation}

Next, we observe that, when $n>1$, the initial potential $\psi_0(x)$ is
a homogeneous random function while, for $n<1$, it only has
homogeneous increments. In the former case, the potential has a
translation-invariant correlation function, denoted by $B_{0\psi}(x)$,
the Fourier transform of which is $E_0(k)/k^2$.  In the latter case, it
is simpler to characterize the initial potential by its structure
function 
\begin{equation}
S_{0\psi}(x) \equiv\langle \left(\psi_0(x)-\psi_0(0)\right)^2\rangle,
\label{defstr}
\end{equation}
which is still translation-invariant
and exists as long as $n>-1$. It is related to the energy
spectrum by
\begin{equation}
S_{0\psi}(x) = 4\alpha^2\int_{0}^{\infty}(1-\cos(kx))\,k^{n-2}\,b_0(k)\,dk.
\label{spsiek}
\end{equation}
When $n>1$ the structure function goes to a finite limit at large
separations while, when $n<1$, it grows without bound\,:
\begin{equation}
   S_{0\psi}(x) \sim \left\{
               \begin{array}{rl}
                \sigma_{\psi}^2    &n>1\\
                \alpha^2|x|^{1-n}&n<1, \\
   \end{array}
   \right. \qquad |x|\to\infty.
   \label{asymptotics}
\end{equation}

An important consequence of (\ref{BEP}), (\ref{vpsi}) and
(\ref{defet}), valid whenever the velocity is homogeneous,
irrespective of the potential being or not being homogeneous, is
\begin{equation}
E(t) = \langle v^2\rangle =  \langle \left(\partial_x \psi\right)^2\rangle =
2 \partial_t \langle\psi(x,t)\rangle.
\label{energyevol}
\end{equation}
It relates the energy to the time-derivative of the mean potential,
which does not dependent on the position $x$.

We now recall some results about the Burgers equations which are relatively
standard and will be used in what follows.

When $n\ge 0$, the correlation invariant
\begin{equation}
J(t) = \int_{- \infty}^{\infty} \langle v(x,t)v(0,t)\rangle \, dx = E(0,t)
\label{correlator}
\end{equation}
is time independent. Indeed, using the conservation form of the
Burgers equation (\ref{BEC}), it is checked that the time derivative
of the integral in (\ref{correlator}), limited from $-X$ to $+X$ is
expressed in terms of correlations between quantities evaluated at
points a distance $X$ apart. Such correlations vanish as $X\to\infty$
because, with our assumptions, the initial conditions have decreasing
correlations at large separation and this property will be
preserved.\footnote{This is expected because influence cannot
propagate faster than the initial velocities. So, correlations over a
distance $X$ should not exceed a value $\propto
e^{-(1/2)X^2/(t\sigma_v)^2)}$.  It should be possible to prove this
rigorously for the kind of initial conditions assumed here.} Note that
$J$ is just the value of the energy  spectrum at $k=0$.

A straightforward consequence of the Burgers equation with a homogeneous
velocity field is
\begin{equation}
\partial_t B_v(x,t) = \frac{1}{6} \partial_x S_3(x,t) +
2 \nu \partial_x^2 B_v(x,t),
\label{KarmanHowarth}
\end{equation}
which relates the correlation function $B_v(x,t)$ and the third order
structure function $S_3(x) \equiv \langle [v(x,t)-v(x,0)]^3\rangle$.
This equation is the analogue of the K\'arm\'an--Howarth equation for
Navier--Stokes turbulence. Just as the latter implies Kolmogorov's
four-fifths law (see, e.g., Frisch 1995),
eq.~(\ref{KarmanHowarth}) implies a ``twelve law''\,: in the limit of
vanishing viscosity and at scales much smaller than the integral
scale, we have
\begin{equation}
S_3(x) = - 12 \varepsilon x,
\label{Law12}
\end{equation}
where $\varepsilon \equiv \partial_t\langle v^2/2\rangle$ is the mean
energy dissipation.

Eq.~(\ref{KarmanHowarth}) or (\ref{Law12}), in the limit of zero
viscosity, puts a strong constraint on self-similar decaying solutions
both for Navier--Stokes and Burgers turbulence. By self-similar
decaying solutions, we understand the following\,: when the
velocity $v$ and the space
coordinate $x$ are  rescaled by suitable time-dependent scaling factors  $u(t)$ and 
$L(t)$, the (single-time)
statistical properties of the solution  become independent of
the time. This definition implies in particular the following scalings
for the correlation function and the third-order structure function of
the velocity\,:
\begin{eqnarray}
B_v(x,t) & = & u^2(t) \tilde B_v\left({x\over L(t)}\right)
\label{Bsim}  \\
S_3(x,t) & = & u^3(t) \tilde S_3\left({x\over L(t)}\right),
\label{Ssim}
\end{eqnarray}
with dimensionless time-independent functions $\tilde B_v$ and $\tilde
S_3$.  Substituting (\ref{Bsim})-(\ref{Ssim}) into
(\ref{KarmanHowarth}) for $\nu\to0$ and demanding that $\tilde B_v$
and $\tilde S_3$ be time independent, we obtain two first-order differential
equations relating $u(t)$ and $L(t)$, namely,
\begin{equation}
{\dot u\over u} =\alpha_1 {\dot L\over L}, \qquad \dot L =\alpha_2 u,
\label{utLtodes}
\end{equation}
where the dot denotes time differentiation and $\alpha_1$ and
$\alpha_2$ are two dimensionless undetermined constants. Note that
exactly the same equations are obtained with a self-similarity ansatz
for Navier--Stokes turbulence. Eq.~(\ref{utLtodes}) has power-law
solutions\footnote{In what follows, we shall be concerned with decaying
solutions which satisfy the self-similarity conditions
(\protect\ref{Bsim})-(\protect\ref{Ssim}) and (\protect\ref{utLtodes})
only in an {\em asymptotic\/} sense, for long times and to leading
order.  We shall then see that $u(t)$ and $L(t)$ may take the form of
power laws with logarithmic corrections.}  but, as already pointed out
by Kolmogorov (1941), the exponents in the power laws cannot be
determined solely from the self-similarity ansatz.  Additional
ingredients are  needed.

For what follows it will be convenient to give a precise definition of the
scaling factors.  We choose $u(t)$ to be the r.m.s.\ velocity
\begin{equation}
u(t) \equiv \langle v^2(x,t)\rangle^{1/2},
\label{uvrms}
\end{equation}
and we define the ``integral scale'' $L(t)$ by
taking $\alpha_2=1$ in (\ref{utLtodes}), that is
\begin{equation} 
\dot L(t) = u(t).
\label{uLt}
\end{equation}

Let us finally note that in Fourier space the self-similarity ansatz
(\ref{Bsim}), together with (\ref{uLt}), translates into
\begin{equation}
E(k,t) = \frac{L^3(t)}{t^2} \tilde E(kL(t)),
\label{ssimilarspectrum}
\end{equation}
\begin{equation}
\tilde E(\tilde k) = \frac{1}{2\pi} \int_{-\infty}^{\infty}
\tilde B_v(\tilde x) \exp(ik \tilde x)\, d \tilde x,
\label{dimlesspec}
\end{equation}
where our choice of normalization of the energy as $ u ^2(t)$ imposes
that $\int\tilde E(\tilde k)\, d \tilde k = 1$.

\section{Phenomenology of the decay}
\label{s:pheno}

\subsection{Permanence of large eddies and self-similar decay}
\label{s:PLE}

An important role in the decay at long times is played by the
principle of {\em permanence of large eddies\/} (PLE).  We shall give
two formulations almost -- but not quite -- equivalent, one in
physical space and the other one in Fourier space.

\par\noindent {\bf  PLE\,: physical
space  formulation.}  We assume that the Gaussian initial velocity has a
correlation function with a power-law behavior at large distances\,:
\begin{equation}
B_{v}(x) \simeq \alpha^2 C_n |x|^{-n-1}, \qquad |x|\to\infty,
\label{Largexinit}
\end{equation}
with an exponent $n>-1$. Then, (\ref{Largexinit}) remains true at any
later time, with the same exponent $n$ and the same constant $\alpha$.
\vspace{1mm}
\par\noindent {\bf  PLE\,: Fourier
space  formulation.} We assume that the Gaussian initial velocity has
a spectrum with a power-law behavior at small wavenumbers\,:
\begin{equation}
E(k) \simeq \alpha^2 |k|^n, \qquad |k|\to 0,
\label{Initspecpowerlaw}
\end{equation}
with an exponent $2>n>-1$. Then, (\ref{Initspecpowerlaw}) remains true
at any later time, with the same exponent and constant.

In its physical-space formulation PLE can be supported by the
following argument. From (\ref{Largexinit}), it follows that the
characteristic velocity associated with a large separation $x$ is
\begin{equation}
u_0(x) \sim \alpha |x|^{-{n+1\over2}},
\label{u0x}
\end{equation}
so that the corresponding ``turnover time'', the characteristic time
for significant changes through nonlinearity, is
\begin{equation}
t_0(x) \sim {x\over u_0(x)} \propto |x|^{n+3\over 2},
\label{t0x}
\end{equation}
which grows without bound at large separations. Hence, at any finite
time, eddies of sufficiently large size should remain essentially as
they were initially. This simple argument could be upset if two fluid
particles, initially strongly correlated (say, separated by not more
than one initial integral scale), were to propagate this correlation
to a large separation $x$ in a finite time.\footnote{For the case of
incompressible Navier--Stokes turbulence, pressure effects can
instantaneously affect the large-scale correlations and thereby
invalidate PLE, at least in the anisotropic case (Batchelor \&
Proudman 1956).} However, with Gaussian initial conditions this
occurrence has a very small probability, which can be bounded by
$\propto \exp [-(1/2)x^2/(\sigma_v t)^2]$, as explained in the
footnote following the derivation of the constancy of $J$
(eq.~(\ref{correlator})) in the previous section.  Thus, the initial
correlation function at large separations should remain unaffected to
leading order.

In its Fourier space formulation, PLE has already been proven in the
previous section, for the case of the spectral index $n=0$. When
$n<2$, PLE is a consequence of the observation that the time-rate of
change of the energy spectrum, the so-called transfer function, is
typically proportional to $k^2$ at small wavenumbers, because of the
beating interactions of pairs of Fourier components with nearly
opposite wavevectors. We may also derive the constancy of the
coefficient of $|k|^n$ just from the constancy of the coefficient of
$|x|^{-n-1}$ at large $x$.

There are reasons why the two formulations are not quite
equivalent. First, the $n=0$ result in Fourier space is not an
instance of PLE in physical space. Indeed, when $n$ is an even
integer, (\ref{Largexinit}) implies a non-analytic spectrum at small
$k$ with logarithmic corrections. Second, when $n>2$ and the initial
correlation function has the behavior (\ref{Largexinit}) at large
separations, the energy spectrum at any later time will begin with
an analytic (time-dependent) $k^2$ term with a subleading
non-analytic correction proportional to $|k|^n$ (with a
time-independent coefficient).

We now show how PLE, used as a ``boundary condition'' at large $x$ or
small $k$ in connection  with the self-similarity ansatz
(\ref{Bsim}) and (\ref{uLt}), allows us to derive the law of temporal
variation of the integral scale and the energy. We first give the
argument in physical space. Let us assume that the distance beyond
which PLE applies is within the scope of the self-similarity formula
(\ref{Bsim}). It then immediately follows from (\ref{Bsim}),
(\ref{uLt}) and
(\ref{Largexinit}) that $\tilde B(\tilde x) \propto |\tilde x|^{-n-1}$
and
\begin{equation}
L(t) \sim (\alpha t)^{\frac{2}{3+n}}
\label{LtSS}
\end{equation}
and thus, by (\ref{uLt}),
\begin{equation}
E(t)=  u ^2(t) \sim \alpha^{\frac{4}{3+n}} t^{-\frac{2(n+1)}{3+n}}.
\label{EtSS}
\end{equation}

In Fourier space the self-similarity ansatz (\ref{ssimilarspectrum}),
together with (\ref{Initspecpowerlaw}) gives $\tilde E(\tilde k) \propto
|\tilde k|^n$, at small $k$ and the same relations for integral scale
and the energy as written above. Clearly, this argument cannot be
applied with initial data such that the spectral index $n\ge 2$, since
the later spectrum has now a $k^2$ dependence at small $k$ with a {\em
time-dependent\/} coefficient.

Actually, we shall now show that the laws (\ref{LtSS}) and
(\ref{EtSS}) apply only when $n<1$.

We begin with the case $-1<n<1$ when the initial potential has
homogeneous increments. Many aspects of this case are well understood,
thanks in particular to Burgers' own work (1974; see also Gurbatov
{\it et al.} 1991).  The phenomenology is quite simple.
Increments $\Delta \psi_0(L)$ of the initial potential over a distance
$L= x-a$ can be estimated from the square root of the structure
function $S_{0\psi}(L)$ of the potential (\ref{defstr}). For a given
position $x$, the maximum in (\ref{MAX})-(\ref{Phifunction}) will come
from those $a$\ms1 such that the change in potential is
comparable to the change in the parabolic term
\begin{equation}
\sqrt {S_{0\psi}(L)}\sim  \frac{L^2}{t}.
\label{following}
\end{equation}
At large times $L$ is also large and we can use (\ref{asymptotics}) to
obtain precisely (\ref{LtSS}) and, hence, the energy law (\ref{EtSS}).
Alternatively, we could argue that when $t$ is so large that the
parabolas appearing in (\ref{Phifunction}) have a radius of curvature
much larger than the typical radius of curvature of features in the
initial potential, we can plausibly replace that initial potential by
fractional Brownian motion of exponent $h=(1-n)/2$ (in order to be
consistent with (\ref{asymptotics})). Without loss of generality we
may assume that this fractional Brownian motion starts at the origin
for $x=0$.  This function is then statistically invariant under the
transformation $x \to \lambda x$ and $\psi_0 \to \lambda ^h\psi_0$. It
is then elementary, using (\ref{MAX})-(\ref{Phifunction}), to prove
that a rescaling of the time is (statistically) equivalent to a
suitable rescaling of $x$ distances and of $\psi(x,t)$. This implies
(\ref{LtSS}) and (\ref{EtSS}).\footnote{For details, see Vergassola
{\it et al.} (1994, Section~4) and Avellaneda  E (1995).}

\subsection{Elementary derivation of the Kida law of decay}
\label{s:Kida}

The case $n>1$ when the initial potential is homogeneous is more
difficult. Because of this homogeneity, the structure function becomes
constant at large distances. If we then formally use (\ref{following})
we obtain a $t^{-1}$ law for the energy and a $t^{1/2}$ law for the
integral scale. This is almost, but not quite, the right result.  There
are logarithmic corrections, first discovered by Kida (1979), which
can actually be captured by phenomenology, as now explained.

We shall estimate the energy in terms of the mean potential at $x=0$,
using (\ref{energyevol}) and the maximum representation
(\ref{MAX})-(\ref{Phifunction}). We assume that, at long  times $t$,
the distribution of values of the potential is peaked around its
(non-vanishing) mean value $\langle \psi\rangle\equiv \langle
\psi(0,t)\rangle$. This quantity is thus the typical maximum value of
$\psi_0(a)-a ^2/(2t)$. Equivalently, we ask\,: what is the value
$\langle \psi\rangle$ by which we should vertically shift the parabola
$a ^2/(2t)$ so that the graph of $\psi_0(a)$ has a substantial
probability of being everywhere below the shifted parabola\,? To
estimate this probability we divide the $a$-axis into equal intervals,
$I_1$, $I_2$, $I_3$, \ldots, with a common length of about one initial
integral scale $L_0$. The initial potential takes approximately
independent values in the different intervals. Thus, the probability of
no intersection with the shifted parabola is (approximately) the
product of the probabilities $q_1$, $q_2$, \ldots of no intersections in the
various intervals\,:
\begin{equation}
P_{\rm no\,\,inters.} \sim q_1 q_2\ldots \,.
\label{Pnointers}
\end{equation}
Each of the
$q_i$\ms1 is of the form $1-p_i$, where $p_i$ is the probability of at
least one intersection in the interval $I_i$. This can be estimated as
\begin{equation}
p_i \sim W\left(\langle \psi\rangle +{a_i ^2\over2t}\right).
\label{piest}
\end{equation}
Here, $W(\psi)$ is the single-point cumulative
probability of $\psi_0$, integrated from $\psi$ to infinity and $a_i$ is
any point in the interval $I_i$, since the parabolic function changes very
slowly with $a$ when $t$ is large. All the $p_i$\ms1 are small and thus
\begin{equation}
P_{\rm no\,\,inters.} \sim 1-p_1 -p_2 \ldots\,.
\label{Pnoqi}
\end{equation}
The sum of the $p_i$\ms1 can be approximated by an integral from
$-\infty$ to $+\infty$. When $\langle \psi\rangle $ is taken very
large this integral is very small. As we decrease $\langle
\psi\rangle$, the probability of having an intersection somewhere
becomes significant when $\langle \psi\rangle $ is chosen such that
\begin{equation}
\int_{-\infty}^{\infty} W\left(\langle \psi\rangle + {a ^2\over
2t}\right)\,{da \over L_0} = O(1).
\label{intorder1}
\end{equation}

Since we assumed Gaussian initial conditions with r.m.s.\ value
$\sigma_{\psi}$,the cumulative probability $W(\psi)$ is approximately
$\exp\left[-(1/2)(\psi/\sigma_{\psi})^2\right]$. The leading order of the
integral in (\ref{intorder1}) is obtained by keeping only the first
two terms in the expansion of the square of $\langle \psi\rangle + a^2/(2t)$.
Thereby, we obtain
\begin{equation}
\langle \psi\rangle \sim
\sigma_{\psi}\ln^{1/2}\left({t\over \tnl}\right),
\label{meanpsipheno}
\end{equation}
where
\begin{equation}
\tnl\equiv {L_0^2\over \sigma_{\psi}}
\label{deftnl}
\end{equation}
is the characteristic nonlinear time.
Using (\ref{energyevol}), we then obtain the Kida (1979) log-corrected
$1/t$ law for the energy decay\,:
\begin{equation}
E(t)\sim t^{-1} \sigma_{\psi}
\ln^{-1/2}\left({t\over \tnl}\right).
\label{Etphenokida}\\
\end{equation}
Finally, assuming (\ref{uLt}), we  obtain Kida's (1979) result for the
integral scale\,:
\begin{equation}
L(t)\sim t^{1/2}(\sigma_{\psi})^{1/2}\ln^{-1/4}\left({t\over \tnl}\right).
\label{Ltphenokida}
\end{equation}
We must, of course, stress that this derivation is only heuristic. We
shall come back to the issue of a more systematic derivation
(Sections~\ref{s:homogeneous} and~\ref{s:correlations}), including
estimates of the subleading corrections. 

It is straightforward to extend this phenomenology to initial
potentials which are homogeneous and non-Gaussian, as the only
ingredient used is the p.d.f.\ of $\psi_0$. For example, when the p.d.f.\ has
a tail $\propto\exp(-|\psi|^\beta)$, the logarithmic correction to the
$1/t$ law has the exponent $(1-\beta)/\beta$.  Hence, when $n>1$ the
law of decay could be  very sensitive to the functional form of the p.d.f.\ of
the initial potential. Such heuristic considerations for non-Gaussian 
potentials have not yet been tested by more systematic theory and simulations.

\subsection{The Kida law and the issue of self-similarity when
$1<n<2$}
\label{s:Kidavsss}

One of the most questionable aspects of our phenomenological
derivation is its validity when $n$ is not an even integer. Indeed,
this gives rise to long-range algebraically decreasing correlations,
which are neglected in this phenomenology and which Kida (1979)
explicitly ruled out.  Actually, we shall see in
Section~\ref{s:homogeneous} that the Kida law remains valid
nevertheless as long as the initial potential is homogeneous, that is
for any $n>1$.
\begin{figure}
\iffigs
\centerline{\psfig{file=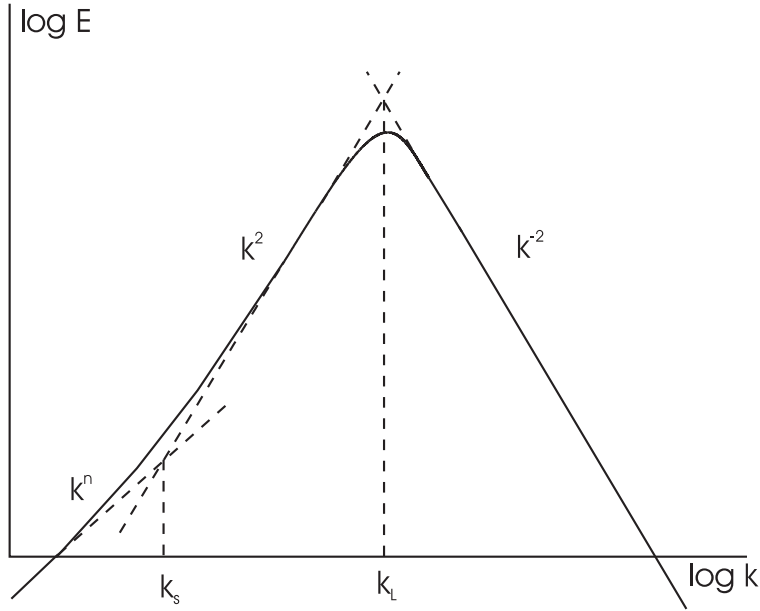,width=10cm,clip=}}
\fi \caption{
Sketch of the energy spectrum at long times when
$1<n<2$.}
\label{f:threeregions}
\end{figure}

We now address for the first time one of the central issues of this
paper.  The predictions (\ref{LtSS}) and (\ref{EtSS}) for $E(t)$ and
$L(t)$, obtained by assuming self-similarity and using PLE as a
``boundary condition'' at large $x$ (or small $k$) are not consistent
with the Kida law (\ref{Etphenokida})-(\ref{Ltphenokida}). Because of
the time-dependence of the coefficient of $k^2$ in the spectrum at
small $k$, the applicability of the former is obviously restricted to
$n<2$. Since the latter applies when $n>1$, there may be a
contradiction in the region of overlap $1<n<2$. Actually, we shall see
that $n>1$ leads to a self-similar decay of the solution in an
``inner'' region of not too large distances (or, in Fourier space, of
not too small wavenumbers), controlled by the integral scale $L(t)$
(\ref{Ltphenokida}). In this inner region correlations of $\psi$
decrease like a Gaussian function and the Fourier signature is a range
of wavenumbers at which the energy spectrum is $\propto k^2$.  In
addition, if $n>1$ and is not an even integer, there exists an
``outer'' frozen region at which PLE applies. Thus, there is, at very
large distances, an algebraic tail in the correlation function and the
spectrum has a time-independent $|k|^n$ contribution at small $k$.
When $1<n<2$ this is the leading term at small $k$ so that the
complete spectrum has three ranges\,: proportional to $|k|^n$ at very
small $k$, then proportional to $k^2$ at intermediate wavenumbers (up
to approximately the inverse of the integral scale $L(t)$) and,
finally, proportional to $k^{-2}$ at high $k$ (the usual signature of
shocks), as shown in Fig.~\ref{f:threeregions}. We shall return to the
issue of the global picture in Section~\ref{s:intermediate}.

\section{Homogeneous initial potential\,: previous work}
\label{s:previouswork}

In this section we discuss previous work dealing with the limit of
vanishing viscosity for initial potentials which are homogeneous and
Gaussian.  The main papers under discussion are Kida (1979), Gurbatov
\& Saichev (1981; see also Gurbatov {\it et al.} 1991),
Fournier \& Frisch (1983) and Molchanov, Surgailis \& Woyczynski
(1995). For initial conditions such that the velocity field is
homogeneous but the potential is not, or such that the initial
conditions are not Gaussian, the reader is referred to Burgers (1974),
Albeverio, Molchanov \& Surgailis {\it et al.}  (1994), and Esipov \&
Newman (1993), Esipov (1994), Hu \& Woyczynski (1996) and references
therein.

Let us first state the essence of what is at presently known to hold
with certainty. In the limit of vanishing viscosity, as the time $t$
tends to infinity, the statistical solution becomes self-similar in
the sense of Section~\ref{s:formulation}. The integral scale $L(t)$
and the energy $E(t)$ are given, to leading order, by
\begin{eqnarray}
L(t)&\simeq& (t\sigma_{\psi})^{1/2}\ln^{-1/4}\left({t\over
2\pi\tnl}\right),
\label{Ltasymptotic}\\
E(t)&\simeq& t^{-1} \sigma_{\psi}
\ln^{-1/2}\left({t\over 2\pi\tnl}\right),
\label{Etasymptotic}
\end{eqnarray}
where
\begin{equation}
\tnl\equiv {L_0^2\over \sigma_{\psi}} =  {L_0\over \sigma_v},
\qquad L_0\equiv {\sigma_{\psi} \over \sigma_v}.
\label{deftnL0}
\end{equation}
The non-dimensionalized self-similar correlation function $\tilde
B_v(\tilde x)$, defined by (\ref{Bsim}), which is a function of
$\tilde x =x/L(t)$, is given by
\begin{equation}
\tilde B_v(\tilde x)= {d\over d\tilde x}\left(\tilde xP(\tilde x)\right),
\label{Bvasymptotic}
\end{equation}
where
\begin{equation}
P(\tilde x) = {1\over2}\int_{-\infty}^\infty {dz \over
g\left({\tilde x+z \over 2}\right) \exp\left[{(\tilde
x+z)^2\over8}\right] + g\left({\tilde x-z \over 2}\right) \exp\left[{(\tilde
x-z)^2\over8}\right]}
\label{defPnoshock}
\end{equation}
and
\begin{equation}
g(z)\equiv \int_{-\infty}^z e ^{-{s^2\over2}}\,ds.
\label{deferror}
\end{equation}
Note that the properties of the self-similar state are universal in so
far as they are expressed solely in terms of two integral
characteristics of the initial spectrum, namely the initial r.m.s.\
potential $\sigma_{\psi}$ and r.m.s.\ velocity $\sigma_v$.\footnote{Observe
that the spectral exponent does not directly enter, in contrast to
what happens when $n<1$ (cf. (\ref{LtSS})-(\ref{EtSS})).} It may be
shown that the function $P(\tilde x)$ is the probability of  having no
shock within an Eulerian interval of length $\tilde x L(t)$
(Gurbatov, Malakhov \& Saichev 1991).

Important consequences of
(\ref{Bvasymptotic})-(\ref{defPnoshock}) are the asymptotic behavior
of the  non-dimensionalized correlation
function $\tilde B_v(\tilde x)$ at large $\tilde x$ and of its Fourier
transform, the non-dimensionalized spectrum $\tilde E(\tilde k)$
(\ref{dimlesspec}) at
small $\tilde k =kL(t)$ (Gurbatov, Malakhov \& Saichev 1991)\,:
\begin{eqnarray}
\tilde B_v(\tilde x) &\simeq&
-\sqrt{\pi\over32}\tilde x\exp\left(-{\tilde x^2\over8}\right),
\qquad |\tilde x|\to \infty,
\label{innerlargex}\\
\tilde E(\tilde k) &\simeq& \mu ^2\tilde k^2, \qquad |\tilde k|\to
0,\label{innersmallk}\\
\mu ^2 &=& {1\over \pi}\int_0^\infty \tilde x^2P(\tilde x)\, d\tilde x
\approx 1.08.
\label{108}
\end{eqnarray}

Most of these results have already been obtained by Kida (1979). He
was interested in the long-time behavior when the initial potential is
homogeneous and the initial correlations decrease rapidly at large
separations, without particularly emphasizing the Gaussian assumption.
He introduced a model of discrete independent maxima whose p.d.f.\ had
three free parameters (an exponent and two constants); their relation
to the properties of the initial conditions (say, the spectrum) were
left unspecified. For the case of a p.d.f.\ with a Gaussian tail, he
obtained the functional form of the results given above.

At long times, the contacts of the parabolas, in the maximum
representation (\ref{MAX})-(\ref{Phifunction}) of the solution, are
near large-value local maxima of the initial potential. Kida's
assumption of independence of the successive maxima implicitly assumes
a Poisson process. Gurbatov \& Saichev (1981) who conjectured the
asymptotic existence of a Poisson process\footnote{See also Fournier
\& Frisch (1983).} and Molchanov, Surgailis
\& Woyczynski (1995) who proved it, showed that, in the $x$-$\psi$
plane, the density of the points is uniform in the $x$-direction and
exponential in the $\psi$-direction. This permits the calculation of
the one- and two-point p.d.f.'s of the velocity (Gurbatov \& Saichev
1981).  Molchanov, Surgailis \& Woyczynski (1995) also calculated the
full $N$-point multiple time distributions. Their proof made crucial
use of some fine properties of extremes of Gaussian processes
(Leadbetter, Lindgren \& Rootzen 1983).\footnote{What they proved is
not exactly in the form of the long-time asymptotic relations given
above but an equivalent result in which the time and space variables
and the amplitude of the initial conditions are rescaled.} From the
point of view of the present paper it is important to stress that
Molchanov, Surgailis \& Woyczynski's (1995) result is valid for any
$n>1$ and not just for $n\ge 2$.\,\footnote{They  assume that
the initial correlations of $\psi$ decrease faster than the inverse of
$\ln |x|$ at large $|x|$ and can be Taylor-expanded to fourth order at
small $|x|$.}

Assuming a Poisson distribution, Gurbatov \& Saichev (1981) and
Fournier \& Frisch (1983) showed that the statistical properties of
the points of contact between the parabolas and the initial potential
can be obtained from the statistical properties of their
intersections, whose mean number can be calculated using the formula
of Rice (1954; see also Leadbetter, Lindgren \& Rootzen 1983; Papoulis
1991). Thus, they could express the parameters in the asymptotic
formulas in terms of the r.m.s.\ initial potential and velocity.

None of the above theories give a control over how fast in time the asymptotic
results are reached. Furthermore, the rescaling involved in these
theories can only capture what we have called in the previous section
the ``inner region'', where PLE is irrelevant. We shall show in the
next section that, for finite long times, the Poisson distribution is
just an approximation which {\em can be systematically
refined}, so as to address such issues.

\section{Systematic derivation of the Kida law for the energy decay 
using the Balian--Schaeffer formula}
\label{s:homogeneous}

The (mean) energy is given by (\ref{energyevol}) in terms of the time
derivative of the mean potential $\left\langle \psi \right\rangle $.
Here, we use the method of Fournier \& Frisch (1983) and the
Balian--Schaeffer formula (\ref{pzerocn}), established in
Appendix~\ref{a:bs}, to derive the asymptotic behavior of the mean
potential for the case of a homogeneous initial potential, that is,
$n>1$.

Let $P(A,t)$ denote the probability density of $\psi $ at time $t$ and
let $Q(A,t)$ denote the (cumulative) probability to have $\psi <A$, given by
\begin{equation}
Q(A,t)=\int\limits_{-\infty }^AP(A^{^{\prime }},t)\,dA^{^{\prime }}.
\label{Intprob}
\end{equation}
The mean value of the potential at time $t$ is then expressible as
\begin{equation}
\langle\psi\rangle =\int\limits_{-\infty }^{\infty}AP(A ,t)\,dA.
 \label{meanvalue1}
 \end{equation}
An integration by parts gives
\begin{equation}
\langle\psi\rangle =\int\limits_{0}^{\infty}[1-Q(A ,t)]\,dA
-\int\limits_{-\infty}^{0}Q(A ,t)\,dA.  \label{meanvalue2}
\end{equation} We shall see that the second integral in
(\ref{meanvalue2}) may be bounded by a time-independent constant,
while the first grows; so it is only the first integral that will be
of importance in what follows.

       From  (\ref{MAX})-(\ref{Phifunction}) and the homogeneity of
the initial potential,  we have
\begin{equation}
Q(A,t)={\rm Prob}\,\left[\psi _0(y)-{y^2\over2t}-A<0,\quad\hbox{for
all}\,\,y
\right].
\label{Intprobpsi}
\end{equation}

\begin{figure}
\iffigs
 \centerline{\psfig{file=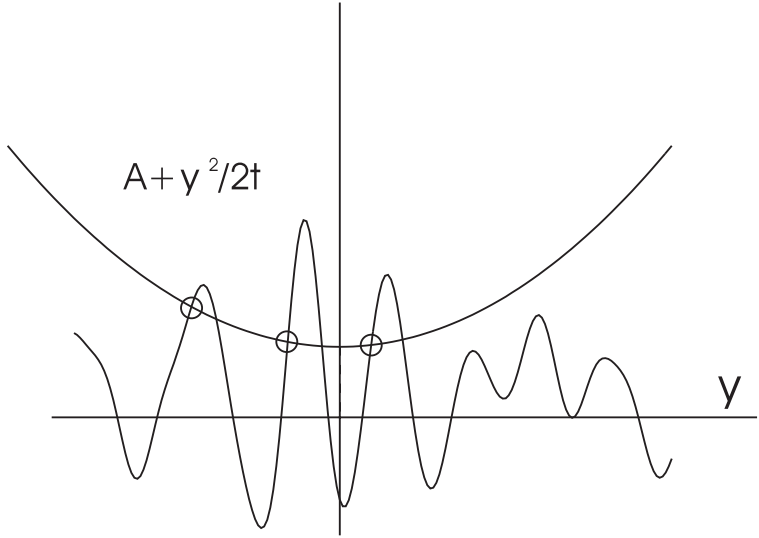,width=10cm,clip=}}
\fi \caption{Intersections of the initial potential $\psi_0(y)$ (curve
$G_1$) with the parabola $G_2$.}
\label{f:one-parabola}
\end{figure}

The argument of the probability in (\ref{Intprobpsi})
can be written as the {\em condition of non-intersection\/} of the
following  two graphs, shown in Fig.~\ref{f:one-parabola}\,:
\begin{eqnarray}
&&G_1:\,\, y\mapsto \psi_0(y), \nonumber\\
&&G_2:\,\, y\mapsto A + \frac{y^2}{2t}.
\label{G1G2intersect}
\end{eqnarray}

We can immediately use this geometric picture to eliminate the second
integral in (\ref{meanvalue2}) from further discussion. Indeed, the
probability that $G_1$ and $G_2$ never intersect is obviously less
than the probability that $\psi_0(y)$ is less than $A$ at the single
point $y$ equal to zero. The integral over $A$ from minus infinity to
zero of this quantity is finite and bounded by a quantity of the order
of $\sigma_{\psi}$. This follows from the observation that, for
negative $A$, the probability $Q(A,t)$ given by (\ref{Intprobpsi}) is
less than the same expression evaluated at the single point
$y=0$. Hence, the contribution of the second integral in
(\ref{meanvalue2}) is bounded by a time-independent constant and will
not contribute to the logarithmically growing leading-order mean
potential.

We now treat the intersections of the graphs $G_1$ and $G_2$ as
``particles'' and apply the Balian--Schaeffer formula (\ref{pzerocn})
of Appendix~\ref{a:bs} to calculate the probability of having zero
particles (no intersections at all). We shall limit ourselves to the
case where $G_1$ intersects
$G_2$ going upward (see the footnote on p.~\pageref{p:upanddown} for
what happens when all intersections are counted.). Such intersections
are depicted by circles in Fig.~\ref{f:one-parabola}.

 From (\ref{Intprobpsi}), (\ref{G1G2intersect}) and 
(\ref{pzerocn}), we have
\begin{equation}
Q(A,t)= p_0 = \exp\left(\sum_{n=1}^\infty {(-)^n\over n!}c_n\right),
\label{IntprobBS}
\end{equation}
where the cumulants $c_n$ are constructed from the $n$-particle
distribution functions $F_n$ as explained in Appendix~\ref{a:bs}. 

In particular, $c_1=\langle N\rangle=\int F_1(y)dy$, the mean number of
particles on the whole real line. We now claim that we have the Rice
(1954) formula\,:
\begin{equation}
\left\langle N\right\rangle =\int F_1(y)dy=\left\langle
\int
\delta
\left(
\Phi \left( y,t \right)
\right)
\{\Phi_y^{_{}^{\prime }}\}^+
dy\right\rangle,
\label{meanN}
\end{equation}
where $\Phi ( y,t)    = \psi_0(y) -\frac{y^2}{2t}-A$ and 
the shorthand $\{x\}^+$ denotes the positive part of $x$,
i.e.\ equal to $x$ if $x$ is greater than zero and zero otherwise and 
$\delta$ is the delta-function.

Indeed, the integral in (\ref{meanN}) only gets a contribution when
the function $\Phi$ vanishes. This happens precisely when the two
graphs $G_1$ and $G_2$ intersect. Let $y^*$ be a point of
intersection.  An integral over a small interval around the
intersection then contributes to (\ref{meanN}) a quantity
$|\Phi_y^{\prime}(y^*,t) |^{-1}\{\Phi_y^{\prime}(y^*,t)\}^+$ .  If the
derivative of $\Phi$ at $y$ is positive, that is, if $G_1$ intersects
$G_2$ going upward, then the contribution is one. If, on the other
hand, $G_1$ intersects $G_2$ going downward, then the contribution is
zero, because $\{\Phi_y^{\prime}(y^*,t)\}^+$ vanishes.  The integral
in (\ref{meanN}) therefore counts the number of upward intersections
(upcrossings) between the two graphs. Note that the assumed smoothness
of the process $\psi_0(y)$ (which follows from the fast decrease at
infinity of its spectrum) renders the applicability of the Rice
formula robust.

The integral $\int F_1(y)dy$ is thus determined by the joint
probability density  $W_{\Phi ,\Phi ^{^{\prime }}}\left( \Phi
,\Phi ^{^{\prime }};y\right) $ of $ \Phi $ and its derivative
$\Phi ^{^{\prime }}$ at the point $y$\,:
\begin{equation}
\left\langle N\right\rangle =\int F_1(y)dy=\int dy\
\int d\Phi ^{^{\prime }}\{ \Phi ^{^{\prime }}\}^+
W_{\Phi ,\Phi ^{^{\prime }}}\left( \Phi ,\Phi
^{^{\prime }};y\right). \label{meanN2}
\end{equation}

We have  similar expressions for the
functions $F_n\left(
y_1,\ldots,y_n\right) $.
Let us introduce the $n$-point
joint probability density  $W_{\psi ,v}\left(
\psi_1,\ldots,\psi_n;v_1,\ldots,v_n \right)$ of the initial potential $\psi$
and the initial velocity $v$, the latter being equal to minus the space derivative
of the initial potential.

Then we can write, analogously to (\ref{meanN2})\,:
\begin{equation}
\begin{array}{c}
\ds F_n\left( y_1,\ldots,y_n\right) =\int dv_1\ldots
dv_nW_{\psi
,v}\left(
\frac{y_1^2}{2t}+A,\ldots
,\frac{y_n^2}{2t}+A;v_1,\ldots,v_n\right)
\cdot \\[1.6ex]
\ds \cdot \{v_1-y_1/t\}^+ \{ v_2-y_2/t\}^+\cdots \{
v_n-y_n/t\}^+.
\end{array}
\label{F1int}
\end{equation}

Since the initial potential is Gaussian, all the joint probability
densities are completely determined by the initial two-point
correlation function\,:
\begin{equation}
B_{0 \psi} \left( x\right)
=\left\langle \psi_0 \left( y+x\right) \psi_0 \left( y\right) \right\rangle.
\label{B0psidef}
\end{equation}
In our case we need the joint probability densities
for (initial) fields  $\psi$ at various points and their (negative) space
derivatives $v$ at those same points.
They are given by 
\begin{equation}
\begin{array}{c}
\ds W_{\psi ,v}\left( \psi _1,\ldots,\psi_n;
v_1,\ldots,v_n\right)
=\frac 1{\left[ \left( 2\pi \right) ^{2n}\det M\right] ^{1/2}}\cdot
\\[1.6ex]
 \ds \cdot \exp \left[-\frac 12\left(
\psi_1^{},\ldots,\psi_n,v_1,\ldots,v_n\right) M^{-1}\left(
\psi_1,\ldots,\psi_n,v_1,\ldots,v_n\right) ^T\right] ,  \end{array}
\label{Wpsiv}
\end{equation}
where $M$ is the enlarged correlation matrix
\begin{equation}
\begin{array}{c}
\ds M_{\psi_\ell ,\psi_{\ell ^{^{\prime }}}}=B_{0 \psi} \left( y_{\ell
}-y_{\ell^{\prime}}\right) \\[1.6ex]
\ds M_{\psi_\ell ,v_{\ell^{\prime }}}=
- M_{v_{\ell^{\prime }},\psi_\ell }=
B_{0 \psi}^{\prime}
\left( y_{\ell}-y_{\ell^{\prime}}\right) \\[1.6ex]
\ds M_{v_\ell ,v_{\ell ^\prime }}= -B_{0 \psi}^{\prime\prime}
\left( y_{\ell}-y_{\ell^{\prime}}\right) =
B_{0v}\left(y_{\ell}-y_{\ell^{\prime}}\right).
\end{array}
\label{defProbpsiv}
\end{equation}
In the last line we have introduced the initial  velocity correlation $B_{0v}$.

The diagonal elements of this matrix are
\begin{equation}
\begin{array}{c}
\ds M_{\psi_{\ell} ,\psi_{\ell} }=\sigma_{\psi}^2, \\[1.6ex]
\ds M_{v_{\ell},v_{\ell} }=\sigma_v^2.
\end{array}
\label{Mdiagelem}
\end{equation}
Here,  $\sigma_{\psi}^2$ and $\sigma_v^2$ are the
variances of the initial potential and the initial velocity,
respectively, as previously defined.

We now return to the Balian--Schaeffer expression (\ref{IntprobBS})
for the cumulative probability of the potential. Inspection of
(\ref{meanvalue2}) shows that the mean potential comes predominantly
from those arguments $A$ such that $Q(A,t)$ is not too close to unity
or, by (\ref{IntprobBS}), such that the alternating sum of
cumulants is not much smaller than unity. As $A$ grows, the changeover
of the alternating sum to a value of order unity happens around a
value which we shall denote $A^*(t)$.

\subsection{The Poisson approximation}
\label{s:Poisson}

We are interested in  the energy decay at
large times and it will be shown that in this case the first
cumulant $c_1=f_1$ is the leading term in (\ref{IntprobBS}) and
that, to leading order, we have the {\em Poisson approximation}\,:
\begin{equation}
Q\left( A,t\right) \simeq
\exp \left\{ -c_1\right\}. \label{IntprobP}
\end{equation}
It is so called for the natural reason that if the upcrossings
obeyed a Poisson distribution, then (\ref{IntprobP}) would be true
exactly. The rest of this subsection contains a computation of the
first cumulant $c_1$.

The function $F_1$ is determined by the single-point probability
density of the initial potential $\psi $ and the velocity $v$, the
subscript $0$ being dropped for conciseness. For a
homogeneous random function, these two variables are 
uncorrelated. We then obtain from (\ref{meanN2}) and (\ref{F1int})
\begin{equation}
F_1\left( y_1\right) =\frac 1{\left( 2\pi \sigma _v^2
\sigma_{\psi}^2\right) }
\int \exp \left( - \frac{1}{2} \left[
\frac{\left( y_1^2/2t+A\right) ^2}{\sigma_{\psi}^2}
+ \frac{v^2}{\sigma _v^2} \right] \right) \cdot
\{ v-\frac{y_1}{t} \}^+ \,dv.
\label{onepointP}
\end{equation}

It is now useful to introduce the auxiliary function
\begin{equation}
Z\left( x\right) =
\left(\frac{1}{2\pi}\right)^{1/2}
\int \{ v-x\}^+ e^{-\frac{v^2}{2}}dv,
\label{Zfunction}
\end{equation}
which can obviously  be expressed in terms of the error
function. We only need to use that, for $x\to 0$,  the function
tends to $Z(0)=\left( 2\pi
\right)^{-1/2}$ and that, for $x\to \infty$, we have   $Z(x)\simeq x$.

In terms of $Z(x)$ the mean number of upcrossings is given by
\begin{equation}
c_1=\int F_1\left( y_1\right) dy_1=\int dy_1\,\sigma_v\,
Z\left(\frac{y_1}{t\sigma_v}\right) \sqrt{1/2\pi
\sigma_{\psi}^2}
\exp \left\{ -\frac{\left( y_1^2/2t+A\right) ^2}
{2\sigma_{\psi}^2}\right\}.
\label{cumulant1}
\end{equation}
The cumulant $c_1$ is a dimensionless number. It is therefore
preferable to rewrite (\ref{cumulant1}) in a manifestly dimensionless
form.  By (\ref{defL0}), $\sigma_{\psi}/\sigma_v$ is equal to $L_0$,
the initial integral scale. In (\ref{cumulant1}) we introduce the new
dimensionless variables $z_1=y_1/L_0 $ and $a=A/\sigma_{\psi}$ and the
dimensionless nonlinear time $\tnl=L_0/\sigma_v$ already defined
(\ref{deftnL0}).  Eq.~(\ref{cumulant1}) can then be written
\begin{equation}
c_1=\int dz_1
Z\left(\frac{z_1}{t/\tnl}\right) \sqrt{1/2\pi}
\exp \left\{ -\frac{\left(
\frac{z_1^2}{2(t/\tnl)}+a\right)^2}{2}\right\}.
\label{cumulant1nondimensional}
\end{equation}
It is seen that (\ref{cumulant1nondimensional}) involves two
spatial scales\,: when $z_1$ is small compared to the ``prefactor
scale'' $t/\tnl$ we 
can simply substitute  $Z(0)$ for $Z(z_1/(t/\tnl))$; when $z_1$ is
large compared to the ``Gaussian scale'' $l(a,t)
\simeq \sqrt{(t/\tnl)/a}$ there is a large damping  by the term in
the exponent quadratic in $z_1$.  If we assume, for the moment, that the
typical value of $a^*=A^*/\sigma_\psi$ where $c_1$ becomes order unity
does not grow in time faster than $\sqrt{t/\tnl}$,\footnote{Actually,
it will be seen to grow only as a square root of the logarithm of the
time.} it follows that it is self-consistent to assume that the
prefactor scale is much larger than the Gaussian scale, so that we may
safely use $Z(0)$ in (\ref{cumulant1nondimensional}).

The remaining integral in (\ref{cumulant1nondimensional}) cannot be
carried out explicitly, but it is nevertheless easy to evaluate it to
relevant leading order. We have two terms in the exponent that depend
on $z_1$\,: one quadratic and one quartic.  The quartic term is larger
for $z_1$ much beyond the Gaussian scale, but then the damping is
strong anyway; so, the correction from the quartic term is small, of
relative order $a^{-2}$ and thus negligible. It is then easily shown
that, for those large values of $a\sim a^*$ which give the dominant
contribution to the mean potential at large-time, we
can use the following leading-order approximation for the cumulant
$c_1$\,:
\begin{equation}
c_1(A,t) \simeq \left(
\frac{t}{\tnl 2\pi a }\right)^{1/2}
e^{-a^2/2}, \qquad
A = \sigma_{\psi}a.
\label{cumulant1A}
\end{equation}
We now determine $a^*(t)$. For this, we assume that $c_1$ is the
dominant term in the Balian--Schaeffer formula (\ref{IntprobBS}) for
the probability of no intersections, as will be shown in
Section~\ref{s:corrections}.  $a^*(t)$ is then determined by the
condition that $c_1(\sigma_{\psi}a ^*,t)$ be order unity. To avoid
undetermined constants, we define $a^*(t)$ by the condition
$c_1(\sigma_{\psi}a ^*,t)=1$, that is through the transcendental
equation
\begin{equation}
\left(\frac{t}{\tnl 2\pi a ^* }\right)^{1/2}
e^{-(a ^*)^2/2} =1,
\label{Trancendentalequation}
\end{equation}
from which it is easily shown that, to leading order, at large times
\begin{equation}
a^* \simeq
\ln^{1/2}\left(\frac{t}{\tnl 2\pi}\right).
\label{Trancesolleading}
\end{equation}

We now return to the cumulative probability of $Q(A,t)$. Using the
Poisson approximation (\ref{IntprobP}) and the leading-order
approximation (\ref{cumulant1A}) for $c_1$, we obtain the following
doubly exponential distribution\,:\footnote{This doubly exponential
distribution is equivalent to the distribution of maxima being Poisson
with uniform density in space and exponentially decaying in $a$
(Molchanov, Surgailis \& Woyczynski 1995).}
\begin{equation}
Q(A,t ) \simeq \exp\{ - \exp \{ -\zeta \} \},
\label{doubleexponential}
\end{equation}
where $\zeta$ is obtained from $a$ by shifting and rescaling\,:
\begin{equation}
a = a^*
\left(1+\frac{\zeta}{\left(a^*\right)^2}\right).
\label{Avarchangea}
\end{equation}
Since $a^*(t)$ is logarithmically large at large times, it follows
from (\ref{doubleexponential})-(\ref{Avarchangea}) that the values of
$a=A/\sigma_\psi$ contributing to the mean potential
(\ref{meanvalue2}) are concentrated near $a^*(t)$ with a small relative
width $\frac{\Delta A}{A^*} = \frac{\Delta a}{a^*} \simeq
\frac{1}{(a^*)^2} \simeq \ln\frac{t}{2\pi\tnl }$. Hence, to leading
order, the mean potential is given in the Poisson approximation by\,:
\begin{equation}
\langle \psi(t)\rangle \simeq \sigma_{\psi}a ^* \simeq \sigma_{\psi}
\ln^{1/2}\left(\frac{t}{2\pi\tnl }\right),
\label{Poissonmeanpsi}
\end{equation}
which implies the energy decay law (\ref{Etasymptotic}).

Notice that, within the Poisson approximation, the energy decay is
fully determined by the local properties of the initial potential
correlation function $B_{0 \psi} \left( x\right) $, namely, $
\sigma_{\psi}^2=B_{0 \psi} \left( 0\right)$ and $\sigma _v^2= - B_{0
\psi}^{^{\prime \prime }}\left( 0\right)$ and does not involve its
large-scale behavior.

\subsection{Corrections to the Poisson approximation}
\label{s:corrections}

So far we have only reproduced a known result. The Balian--Schaeffer
formula allows us to find the applicability condition for the Poisson
approximation by estimating contributions of higher-order cumulants to
(\ref{IntprobBS}).

We thus turn to the calculation  of the second cumulant $c_2$ appearing in
(\ref{IntprobBS}), given in terms of the one- and  two-particle
densities by (\ref{c1c2f1f2}), repeated hereafter for convenience\,:
\begin{equation}
c_2=\int\int
\left[
F_2( y_1,y_2) -F_1( y_1)F_1( y_2)
\right]
dy_1 dy_2.
\label{c2F2F1}
\end{equation}
The function $F_2$ is now given by (\ref{F1int}).  Our purpose in this
section will only be to show that the contribution of $c_2$ -- and of
the higher-order $c_n$\ms1 -- are subdominant compared to that of
$c_1$. For this purpose, it will be enough to estimate the order of
magnitude of $c_2$ (for values of $a\sim a^*(t)$). Thus, there is no
need to derive the full leading-order expression for $c_2$ with all the
numerical constants.

 From (\ref{F1int})-(\ref{defProbpsiv}) it follows that, unlike $c_1$,
the second cumulant $c_2$ involves the entire initial correlation
function $B_{0 \psi}(x)$. Let us denote by $\Delta_{\hbox{\small
corr}}$ a typical correlation length for the initial
potential.\footnote{We could have taken $\Delta_{\hbox{\small corr}}
=L_0$ but keeping the distinction has some interest, as we shall see.}
We shall now distinguish those contributions to (\ref{c2F2F1}) where
the two arguments $y_1$ and $y_2$ are separated by not more than
$\Delta_{\hbox{\small corr}}$ and those where the separation is larger
or, possibly, much larger, a case of interest when the initial
correlation function has a slow algebraic decrease. The former will be
called ``local'' (loc.) and the latter ``non-local'' (nloc.).

Using (\ref{F1int}) and (\ref{c2F2F1}) we can estimate 
\begin{equation}
c_2^{\hbox{\small loc.}} \sim
\int\int F_2\left( y_1,y_2\right) dy_1dy_2
\sim
\Delta_{\hbox{\small corr}}\int \left(F_1( y_1)\right)^2 dy_1.
\label{c2local}
\end{equation}
Introducing as in (\ref{cumulant1nondimensional})
new (dimensionless) variables
$z_1$ and $a$, we get
\begin{equation}
c_2^{\hbox{\small loc.}} \sim
\frac{\Delta_{\hbox{\small corr}}}{L_0}
\int dz_1
Z\left(\frac{z_1}{t/\tnl }\right)^2 \frac{1}{2\pi}
\exp \left\{ -\left(
\frac{z_1^2}{2(t/\tnl )}+a\right)^2\right\}.
\label{c2localnondimensional}
\end{equation}
The integral in (\ref{c2localnondimensional})
is practically the same as for the first cumulant,
so we have
\begin{equation}
c_2^{\hbox{\small loc.}}(A,t) \sim
\frac{\Delta_{\hbox{\small corr}}}{L_0} \left(
\frac{t}{\tnl  a }\right)^{1/2}
e^{-a^2}, \qquad
A = \sigma_{\psi}a.
\label{c2localresult} \end{equation}
The main differences with the expression (\ref{cumulant1A}) for $c_1$
is (i) the presence of a factor $\Delta_{\hbox{\small corr}}/L_0$ and (ii) the
fact that the argument of the exponential is now $-a ^2$ instead of $-a ^2/2$.
 From (\ref{Trancendentalequation}) and (\ref{c2localresult}) we find
that, around the value $a^*(t)$ which give the dominant contribution
to the mean potential, the second order local cumulant is roughly
\begin{equation}
c^{\hbox{\small loc.}}_2 \sim
\left( \frac{\Delta_{\hbox{\small corr}}}{L_0}\right) \cdot \left(
\frac{t}{\tnl  a^*(t)} \right)^{-1/2}.
\label{c2localestimate}
\end{equation}
It easily follows that the local second order cumulant  may
be neglected compared to $c_1$ as soon as 
\begin{equation}
t \gg \tnl 
\left(\frac{\Delta_{\hbox{\small corr}}}{L_0}\right)^2.
\label{c1dominanslocal}
\end{equation}

Here, we make a short digression. There are instances where
$(\Delta_{\hbox{\small corr}}/L_0)^2$ can be large.  Consider for
instance a quasimonochromatic signal with a center wavenumber $k_0\sim
L_0^{-1}$ and a width $\Delta k\sim \left[\Delta_{\hbox{\small
corr}}\right]^{-1} \ll k_0$.  In this case, the condition $t/\tnl \gg
1$ is not enough for the  Poisson approximation
(\ref{IntprobP}) to hold.  The asymptotic regime described by Kida's
law is then obtained much later. As long as $\tnl \ll t \ll \tnl
(\Delta_{\hbox{\small corr}}/L_0)^2$, the energy decays much faster\,:
$E(t) \sim L_0^2/t^2$ (Gurbatov \& Malakhov 1977).  The physical
reason for this is a strong correlation of the shocks in the early
stage of the evolution, which prevents the rapid merging of
shocks.\footnote{Similar remarks can be made when the initial energy spectrum
has a very long plateau at intermediate wavenumbers.}

Let us now turn to the non-local contribution to $c_2$ which is
important only when $n$ is not an even integer, so that there is an
algebraic tail in the correlation function\,: 
\begin{equation}
B_{0 \psi} \left( x\right) \sim \sigma_{\psi}^2 \left({\Delta_{\hbox{\small corr}}
\over x}\right) ^{n-1},\qquad \qquad x\gg \Delta_{\hbox{\small corr}}. 
\label{largexB}
\end{equation}
Owing to this, when $1<n<2$ the main contribution to $c_2$ comes from
the non-local region where $\left| y_1-y_2 \right| \gg L_0$ while $\left|
y_1\right|$, $\left| y_2\right| \leq L(t)$.  Here, $L(t)$ is the
integral scale at time $t$, given by (\ref{Ltphenokida}).

The main technical difficulty now is that we have to express the
inverse matrix of correlations that appears in the exponent in the
Gaussian joint probability densities
(\ref{F1int})-(\ref{defProbpsiv}).  Actually, there is no need to
fully invert the matrix.  It is sufficient to note that when the
points are far apart, correlations become weak and the off-diagonal
elements will be small compared to the diagonal ones.  We may
therefore invert perturbatively to first order in the off-diagonal
elements. Furthermore, the velocity-velocity correlations and the
velocity-potential correlations decay much faster  with separation than the
potential-potential correlations.  Therefore, it is consistent to keep
in the inverted matrix only off-diagonal elements proportional to
$B_{0 \psi}(y_{\ell}-y_{\ell'})$.  As before we can substitute for the
function $Z(x)$ its value at the origin, 
$(2\pi)^{-\frac{1}{2}}$.  After all these simplifications we can write 
the following expression for the non-local part of the second
cumulant $c_2$\,:
\begin{equation}
\begin{array}{c}
\ds c_2^{\hbox{\small nloc.}} = \frac{1}{(2\pi L_0)^2}
\int_{\rm NL} \, dy_1 \, dy_2 \cdot
\exp \left\{ -\frac{1}{2 \sigma_{\psi}^2} \left[ \left(
\frac{y_1^2}2+A\right) ^2+\left( \frac{y_2^2}2+A\right) ^2\right]
\right\} \cdot  \\[2.3ex] 
\ds \cdot \left\{ \exp \left\{ \frac{B_{0 \psi} \left(
y_1-y_2\right) }{\sigma _\psi ^4}\left[ \left(
\frac{y_1^2}2+A\right) \left( \frac{y_2^2}2+A\right) \right] \right\}
-1\right\}.
\end{array}
\label{c2nonlocal1}
\end{equation}
Here, the subscript NL (non-local) is a reminder that we only integrate over
points $y_1$ and $y_2$ sufficiently far apart.  The next step is to
expand the second exponential, which contains the correlation
function, to linear order.  As in the evaluation of $c_1$, we change
the variables $y_1$, $y_2$ and $A$ into the dimensionless variables
$z_1$, $z_2$ and $a$, respectively and obtain, using (\ref{largexB})\,:
\begin{eqnarray}
c_2^{\hbox{\small nloc.}} & \sim &
\left( \frac{\Delta_{\hbox{\small corr}}}{L_0} \right)^{n-1} \cdot
\frac{1}{(2\pi)^2} \cdot \nonumber \\
&& \cdot \int_{\rm NL} \, dz_1 \, dz_2 \cdot
\exp \left\{ - \frac{1}{2} \left(
\frac{z_1^2}{2(t/\tnl)}+a\right)^2+
\left(\frac{z_1^2}{2(t/\tnl)}+a\right)^2\right\}\cdot \nonumber \\
&& \cdot |z_1-z_2|^{1-n}
\left(\frac{z_1^2}{2(t/\tnl)}+a\right)
\left(\frac{z_2^2}{2(t/\tnl)}+a\right),
\label{c2nonlocal}
\end{eqnarray}
the integration being now over the region such that 
$|z_1-z_2|>\Delta_{\hbox{\small corr}}/L_0$.

The integral (\ref{c2nonlocal}) is estimated in the following way\,:
the exponentials effectively delimit the region of integration
to a circle in the $(z_1,z_2)$-plane with radius
$l(t) \sim \left(t/(a\tnl)\right)^{1/2}$.
Inside this circle, most of
the factors in (\ref{c2nonlocal}) are of order unity. Therefore,
we can estimate the non-local part of $c_2$ as
\begin{equation}
c_2^{\hbox{\small nloc.}} \sim 
\left( \frac{\Delta_{\hbox{\small corr}}}{L_0} \right)^{n-1} \cdot
\frac{1}{(2\pi)^2}
\int_{\rm NL}
\, dz_1 \, dz_2 \cdot
e^{-a^2} a^2
|z_1-z_2|^{1-n},
\label{Ansatz}
\end{equation}
where NL is now the region
\begin{equation}      \\
|z_1|< \left( \frac{t}{\tnl a} \right)^{\frac{1}{2}},\qquad
|z_2|< \left( \frac{t}{\tnl a} \right)^{\frac{1}{2}}, \qquad
|z_1 - z_2| > \frac{\Delta_{\hbox{\small corr}}}{L_0}.
\label{NLregion}
\end{equation}

When $n > 2$ the non-local component of $c_2$ is much smaller than the
local one and we may ignore it.  When
$1<n<2$ the integral in (\ref{Ansatz}) is readily evaluated, and we
obtain the estimate
\begin{eqnarray}
c_2^{\hbox{\small nloc.}}& \sim &
\left( \frac{\Delta_{\hbox{\small corr}}}{L_0} \right)^{n-1} \cdot
e^{-a^2} a^2
\left( \frac{t}{\tnl a} \right)^{\frac{1}{2}}
\cdot \left( \frac{t}{\tnl a} \right)^{\frac{2-n}{2}}.
\label{Ansatzresult}
\end{eqnarray}
Finally, estimating $c_2^{\hbox{\small nloc.}}$ for $A\simeq A ^*(t)$,
we obtain
\begin{equation}
c_2^{\hbox{\small nloc.}} \sim
\left( \frac{\Delta_{\hbox{\small corr}}}{L_0}\right)^{n-1}
\left( \frac{\tnl }{t}\right) ^{(n-1)/2}.
\label{c2nlest}
\end{equation}
Comparison with (\ref{c2localestimate}) indicates that the non-local
part of $c_2$ is much larger than the local one, but still small
compared to unity, so that we may indeed neglect $c_2$ compared to
$c_1$.\label{p:upanddown}\footnote{This statement holds provided only
upcrossings or only downcrossings between the graphs $G_1$ and
$G_2$ are counted.  Otherwise, two things happen\,: (i) the mean number
of crossings and hence $c_1$ is doubled; (ii) strong correlations
are present between successive up and down large-ordinate
crossings; these invalidate the Poisson approximation; actually,
the coefficient $c_2$ becomes large and exactly cancels the
aforementioned factor two.}  Let us observe that when $n$ is
just slightly larger than one, an exceedingly long time may be needed
before the asymptotic Kida law applies since  $c_2$ is 
then just marginally smaller than $c_1$.

Using similar procedures we can show that $c_m^{\hbox{\small
loc.}}$ and $c_m^{\hbox{\small nloc.}}$ for $m>2$ are even smaller than
the $c_2$-corrections. This establishes the validity of the Kida law
(\ref{Etasymptotic}) at long times.

\section{The correlation function at long times and the preservation
of long-range correlations}
\label{s:correlations}

In this section we calculate the (spatial) correlation function of the
potential at long times for the solution to the  Burgers equation, using a
Balian--Schaeffer formula, adapted now to the evaluation of two-point
quantities.  As we shall see, in evaluating the far tail of the
correlation function, the dominant contribution will come from the
second rather than from the first cumulant. 

For reasons of symmetry we take the following definition for the
correlation function of the potential\,:
\begin{equation}
B_{\psi}(x,t) = \langle\psi(x/2,t)\psi(-x/2,t)\rangle -
\langle\psi(x,t)\rangle^2,
\label{dBpsi}
\end{equation}
from which the  the velocity correlation function is obtained as
\begin{equation}
B_{v}(x,t) = - \frac{\partial ^2 B_{\psi}(x,t)}{\partial x^2}.
\label{dBv}
\end{equation}

To evaluate (\ref{dBpsi}) we need, in addition to  the one point
probability density $P_1(A,t)$, the joint probability density
$P_2(A,B;x,t)$ of the potential at the same time $t$ for two points
with spatial separation $x$. From (\ref{dBpsi}), we obtain
\begin{equation}
B_{\psi} = \int \int
AB(P_2(A,B;x,t) - P_1(A,t)P_1(B,t))\,dAdB.
\label{B-psi}
\end{equation}
We also introduce the cumulative probability $Q_{2}(A,B;x,t)$, related
to $P_2(A,B;x,t)$ by
\begin{eqnarray}
P_2(A,B;x,t) & = & \frac{\partial^2}{\partial A \partial B}
Q_{2}(A,B;x,t).
\label{Q2relatedtoP2}
\end{eqnarray}
Using this in (\ref{B-psi}) and integrating by parts twice, we have
\begin{equation}
B_{\psi}(x,t) = \int \int_{-\infty}^{\infty}
[Q_{2}(A,B;x,t) - Q_1(A,t)Q_1(B,t)]\,dAdB.
\label{B-psi1}
\end{equation}

\begin{figure}
\iffigs
\centerline{\psfig{file=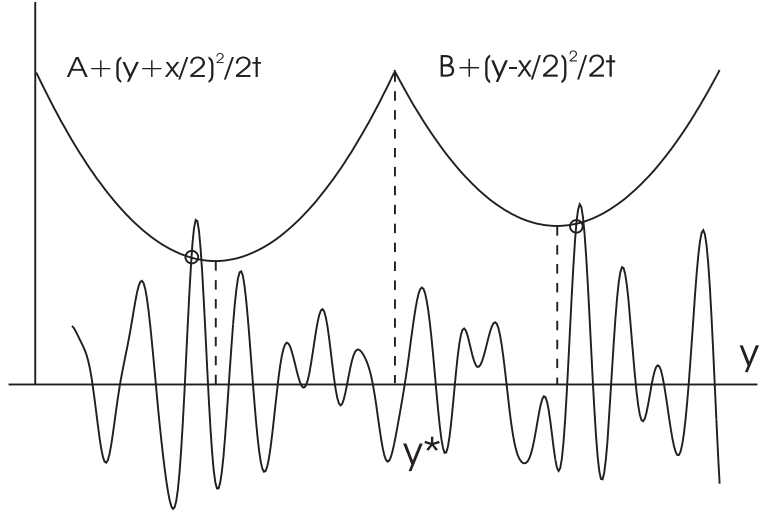,width=10cm,clip=}}
\fi \caption{
Intersections of the initial potential $\psi_0$ with
two parabolas.}
\label{f:two-parabolas}
\end{figure}

By the maximum representation solution (\ref{MAX}) we can express
$Q_{2}$ in the form
\begin{equation}
Q_{2}(A,B;x,t) = 
{\rm Prob}\,\left(G_1\,\, \hbox{never intersects}\,\, G_2\right),
\label{G1neverG2}
\end{equation}
where $G_1$ and $G_2$ are now the following graphs, shown in
Fig.~\ref{f:two-parabolas}\,:
\begin{eqnarray}
&&G_1:\,\, y\mapsto \psi_0(y), \nonumber\\
&&G_2:\,\, y\mapsto g(y)=\min
\left[
\frac{\left(y-\frac{x}{2}\right)^2}{2t}+A,\,
\frac{\left(y+\frac{x}{2}\right)^2}{2t}+B
\right].
\label{G1-G2}
\end{eqnarray}
Thus, the procedure of calculation of $Q_{2}$ is similar to that used
in the preceding section for calculating the mean potential, but with a
more complex function $g(y)$. Hence, we may again use the
Balian--Schaeffer formula
and obtain
\begin{equation}
Q_2(A,B;x,t)=\exp\left(\sum_{n=1}^\infty {(-)^n\over n!}c_n\right),
\label{Q2BS}
\end{equation}
where the $c_n$\ms1 depend now on $A$, $B$, $x$ and $t$.

The first cumulant $c_1(A,B)$ in the expansion (\ref{Q2BS}) is now
\begin{equation}
c_1(A,B) =
\int Z\left(g'(y)/\sigma_v\right)
\frac{1}{(2 \pi \sigma_{\psi})^{1/2}}
\,\exp
\left\{
- \frac{1}{2} \frac{g(y_1)}{2 \sigma_{\psi}^2}
\right\} \, dy_1.
\label{c1expan}
\end{equation}

 From (\ref{G1-G2}) and Fig.~\ref{f:two-parabolas} we see that the
integral in (\ref{c1expan}) can be divided into two integrals
over pieces of the two  parabolas intersecting  at $y ^*$\,:
\begin{eqnarray}
c_1(A,B) & = &
\frac{1}{2 \pi L_0}
\left[
\int_{-\infty}^{y_*}
\exp
\left\{
-\frac{
\left[
\frac{
\left( y + \frac{x}{2} \right)^2
}
{
2t
}
+ B
\right]^2
}
{
2 \sigma_{\psi}^2
}
\right\}
\, dy
\right.
\nonumber \\
& + &
\left.
\int_{y_*}^{-\infty}
\exp
\left\{
-\frac{
\left[
\frac{
\left(
y + \frac{x}{2}
\right)^2
}
{
2t
}
+ A
\right]^2
}
{
2 \sigma_{\psi}^2
}
\right\}
\, dy
\right],
\label{2ints} \\
y_* & = & \frac{A-B}{2x} t.
\label{y*}
\end{eqnarray}
Here, we have assumed from the start that $t \gg \tnl$, to replace $Z(x)$ by 
$Z(0)=(2\pi)^{-1/2}$.

It may now be checked that (\ref{2ints}) behaves in different ways
depending on how the widths of the Gaussians compare to the separation
$x$.

As long as $x$ is smaller than or comparable to the integral scale
$L(t)$ given by (\ref{Ltasymptotic}), the leading-order term in
(\ref{Q2BS}) is found to be $c_1$, that is, the Poisson approximation
remains valid.  One then derives the expressions
(\ref{Bvasymptotic})-(\ref{defPnoshock}) for the correlation function
previously obtained by Kida (1979), Gurbatov \& Saichev (1981) and
Molchanov, Surgailis \& Woyczynski (1995). Actually, to derive the
self-similarly evolving correlation function given by
(\ref{Bvasymptotic}), the only possible spatial scaling factor is (up
to a multiplicative constant) the integral scale $L(t)$ given by
(\ref{Ltasymptotic}). It may also be checked that $L(t)$ is (up to a
constant factor) equal to $\langle \psi^2(t)\rangle^{1/2} /\langle
v^2(t)\rangle^{1/2}$

When $|x|\gg L(t)$ the situation is quite different. Now, the two
parabolas intersect at a very large ordinate so that, to leading
order, we have two independent integrations over
two different almost complete  parabolas.  Thus, to leading order, we
have 
\begin{equation}
c_1(A,B;x,t) \simeq c_1(A,t) + c_1(B,t),\qquad x \gg
L(t).
\label{c1cancellation}
\end{equation}
This, however, implies
\begin{eqnarray}
Q_{2}(A,B;x,t) &-& Q_1(A,t)Q_1(B,t)  \simeq 
\exp\{-c_1(A,t)-c_1(B,t)\} \nonumber \\
& - & \exp\{-c_1(A,t)\}\exp\{-c_1(B,t)\}  = 0,
\label{Q2Poisson}
\end{eqnarray}
which, in turn, implies the vanishing of $B_{\psi}(x,t)$. It follows
that the leading-order contribution cannot come from the Poisson
approximation. 

\begin{figure}
\iffigs
\centerline{\psfig{file=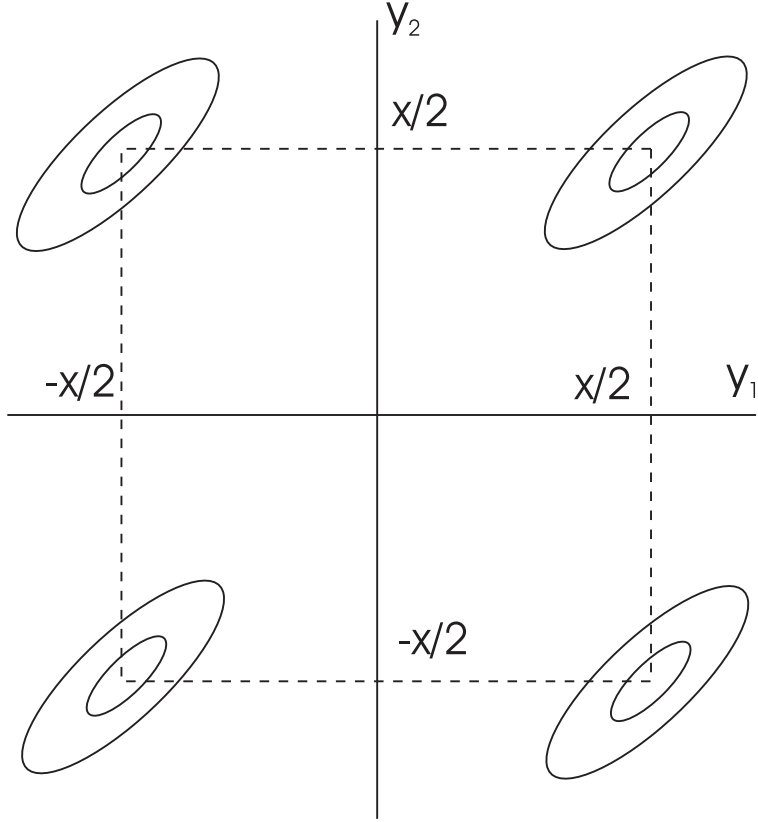,width=10cm,clip=}}
\fi \caption{
The different regions of integration appearing in
(\protect\ref{c2intexpr}). Near the diagonal $y_1=y_2$ we have regions I and I'
of (\protect\ref{IandII}). Away from the diagonal we have regions II
and II'.}
\label{f:regions-of-integration}
\end{figure}

This leading-order contribution actually comes from the second order cumulant
\begin{equation}
c_2 = \int \int [F_2(y_1,y_2) - F_1(y_1)F_1(y_2)] \, dy_1 dy_2 ,
\label{c2intexpr}
\end{equation}
where $F_1$ and $F_2$ are now the one- and two-particle densities
constructed from the crossings of the $G_1$ and $G_2$ graphs defined
in (\ref{G1-G2}).  It may be checked that when $|x| \gg L(t)$ the
significant regions in (\ref{c2intexpr}) are\,:
\begin{equation}
\begin{array}{lr}
\ds\mbox{I}: \; (y_1 \sim \frac{x}{2},\,y_2 \sim \frac{x}{2}), \;& \ds
\mbox{I'}: \; (y_1 \sim -\frac{x}{2},\,y_2 \sim -\frac{x}{2}),\\[1.6ex]
\ds \mbox{II}: \; (y_1 \sim \frac{x}{2},\,y_2 \sim -\frac{x}{2}), \;&\ds
\mbox{II'}: \; (y_1 \sim -\frac{x}{2},\,y_2 \sim \frac{x}{2}),
\end{array}
\label{IandII}
\end{equation}
also illustrated in Fig.~\ref{f:regions-of-integration}.

In the regions I and I' we have the same estimates
(\ref{c2localestimate})-(\ref{c2nlest}) for the value of $c_2$ as in
the calculation of the mean potential. The corresponding contribution
to the correlation function goes to zero with separation even faster
than the $c_1$-contribution and is not relevant.

In the regions II and II', for $t \gg \tnl$ and $|x| \gg L(t)$ we
obtain after integration over $v_1$, $v_2$, the following expression
\begin{eqnarray}
\frac{1}{2} c_2 & = & \frac{1}{(2\pi L_0)^2}
\int \int_{-\infty}^{\infty} \, dy_1 dy_2
\exp
\left\{
- \frac{\left( \frac{y_1^2}{2t} + A \right)^2}{2\sigma_{\psi}^2}
- \frac{\left( \frac{y_2^2}{2t} + B \right)^2}{2\sigma_{\psi}^2}
\right\} \cdot \nonumber \\
&& \cdot
\left(
\exp
\left\{
B_{0\psi}(x)
\frac
{
\left( \frac{y_1^2}{2t} + A \right)
\left( \frac{y_2^2}{2t} + B \right)
}
{
\sigma_{\psi}^4
}
\right\}
- 1
\right) \nonumber \\
& \simeq &
c_1(A)c_1(B)
\left(
\exp
\left\{
\frac{ABB_{0 \psi}(x)}{\sigma_{\psi}^4}
\right\}
- 1
\right).
\label{c2expr}
\end{eqnarray}
Just as in Section~\ref{s:Poisson}, the joint probability density of
$A$, $B$ is  localized in a  narrow region near $A\simeq B \simeq 
\langle \psi\rangle\simeq \sigma_{\psi}(\ln t/2 \pi \tnl)^{1/2}$.
We introduce in (\ref{c2expr}) the new variables $\zeta$ and $\xi$ through
\begin{equation}
A = \langle\psi\rangle  + \frac{\sigma_{\psi}^2}{\langle\psi\rangle }\zeta, \;
B = \langle\psi\rangle  + \frac{\sigma_{\psi}^2}{\langle\psi\rangle }\xi  \;,
\label{newAB}
\end{equation}
and we observe that, when $\zeta $ and $\xi $ are $O(1)$, we have, by
(\ref{cumulant1A}) and (\ref{Poissonmeanpsi}),
\begin{equation}
c_1(A)\simeq e^{-\zeta},\qquad c_1(B)\simeq e^{-\xi}.
\label{c1ABzetaxi}
\end{equation}
We now evaluate the correlation of the potential (\ref{B-psi1}) for
$|x| \gg L(t)$. For this we use (\ref{Q2BS}), truncated to its first
two terms (\ref{c2expr}) and (\ref{c1ABzetaxi}), and the fact that
$B_{0 \psi}(x)$ is small, so that we can expand exponentials with
$B_{0 \psi}(x)$ in their arguments. We finally obtain\,:
\begin{equation}
B_{\psi}(x,t) \simeq B_{0\psi}(x),\qquad t \gg \tnl,\quad |x| \gg L(t).
\label{Bpsiconservation}
\end{equation}

This establishes the preservation of long-range correlations, an
instance of the permanence of large eddies (PLE), as defined 
in Section~\ref{s:pheno}.

\section{The global picture at long times}
\label{s:intermediate}

We are now in a position to derive the main results of this paper,
some of which have  already been presented in a phenomenological way in
Section~\ref{s:pheno}. 

Our derivation of the Kida law (\ref{Etasymptotic}) for the energy
decay in Section~\ref{s:homogeneous}, using the Balian--Schaeffer
formula, proves the validity of this law for any $n>1$ and not just
$n>2$ as thought previously. Similarly, for any $n>1$ the integral
scale $L(t)$ is given by (\ref{Ltasymptotic}).

More novel results emerge when considering the global structure, at
long times, for the correlation function (in physical space) and for its
Fourier transform, the energy spectrum. We shall consider only the
situation at scales larger than the integral scale $L(t)$. Indeed, we
have already stated in Section~\ref{s:correlations} that, at scales
small or comparable to the integral scale, we recover known results
(see Kida 1979; Gurbatov, Malakhov \& Saichev 1991).

We observe that for any $n>1$ which is not an even integer the
correlation function $B_v(x)$ has two different asymptotic regions.
In an ``outer asymptotic region'' at extremely large $x$, permanence of
large eddies (in the form (\ref{Bpsiconservation}) established in
Section~\ref{s:correlations}) implies that 
\begin{equation}
B_{v}(x,t) \simeq B_{v}(x,0) \simeq C_n\alpha^2 |x|^{-n-1}, \qquad
|x|\to\infty.
\label{largexBv}
\end{equation}
There is also an ``inner asymptotic region'' where the correlation
function has the (approximately) Gaussian decrease given by
(\ref{innerlargex}) in dimensionless form with $\tilde x= x/L(t)$.
Using (\ref{Bsim}) we can rewrite this in
dimensioned form as
\begin{equation}
B_v\left( {x\over L(t)}\right) \simeq -\sqrt{\pi\over32}\,
u^2(t)\,{x\over L(t)}\,\exp\left(-{x^2\over8L^2(t)}\right),
\label{innerlargexdim}
\end{equation}
which can be reexpressed in terms solely of $L(t)$, using $u(t)= \dot
L(t)$.  The separation $l_s(t)$ at which the correlation function
switches from its inner to its outer form, called here the {\em
switching distance}, is obtained by equating (\ref{largexBv}) and
(\ref{innerlargexdim}), yielding\footnote{To truly characterize the
transition from the inner to outer asymptotic expansion it is
necessary to perform matched asymptotic expansions which require more
than the leading order.  Equating the leading-order inner and outer
terms correctly predicts the order of magnitude of the scale at which
this transition takes place.}
\begin{equation}
l_s(t) \sim L(t) \left((n-1)\ln \left({t\over\tnl}\right)\right)^{1/2}.
\label{switchX}
\end{equation}
Hence, the switching distance is only logarithmically larger than the
integral scale.\footnote{When $n$ is an even integer the
algebraic outer region disappears altogether.} Still, the fact that
$L(t)$ and $l_s(t)$ have (slightly) different scaling laws in time
implies that, {\em globally, the correlation function does not evolve
in a self-similar fashion.}

This phenomenon becomes much more prominent if we consider the energy
spectrum, the Fourier transform of the correlation function. In
dimensionless inner-region variables, the energy spectrum has the
following form (Gurbatov, Malakhov \& Saichev 1991)\,:
\begin{equation}
\tilde E(\tilde k)
\simeq \left\{ 
\begin{array}{rl}
0.36 {\tilde k}^{-2}, & \tilde k \gg 1 \\[1.5ex]
1.08 {\tilde k}^2,    & \tilde k \ll 1 \\
\end{array}
\right., \quad \tilde k \equiv kL(t).
\label{tEasympt}
\end{equation}
The ${\tilde k}^{-2}$ region is the signature of shocks and will not
interest us further, while the
${\tilde k}^2$ region comes about because the (inner) correlation function
of the potential has a non-vanishing integral from $-\infty$ to
$+\infty$. In dimensioned variables the small-$k$ inner-region behavior
of the spectrum  is thus 
\begin{equation}
E(k,t) \simeq \frac{L^5(t)}{t^2} \,k^2 \simeq  A(t) \,k^2,\quad kL(t)\ll
1, \quad \hbox{inner region},
\label{Eleft}
\end{equation}
where
\begin{equation}
A(t) = \sigma_v^2 L_0^3 \left( \frac{t}{\tnl } \right)^{\frac{1}{2}}
\,
\ln^{-5/4} \left(\frac{t}{2 \pi \tnl }\right).
\label{Aoft}
\end{equation}
Instead of a Gaussian region (in physical space), we have a spectrum
with an algebraic $k^2$ region and a time-dependent coefficient
$A(t)$.

We must now  distinguish two cases.  When $n>2$, the $k^2$ contribution
(\ref{Eleft}) dominates everywhere over the $|k|^n$ contribution, the
latter being then only a subdominant term in the small-$k$
expansion.\footnote{This subdominant term is easily seen in numerical
simulations at very short times and moderately small wavenumbers, for
example when $n=3$ (A.~Noullez, private communication).}  The more
interesting case is $1<n<2$, already sketched in
Fig.~\ref{f:threeregions}.  The permanence of large eddies implies now
that, at extremely small $k$,
\begin{equation}
E(k,t) \simeq \alpha^2 |k|^n,\quad  {\rm for}\,\, k\to 0, 
\quad \hbox{outer region}.
\label{Eksmall}
\end{equation}
This relation holds only in an outer region $|k|\ll k_s(t)$ where
(\ref{Eksmall}) dominates over (\ref{Eleft}). The switching wavenumber
$k_s(t)$, obtained by equating (\ref{Eleft}) and (\ref{Eksmall}), is
given by
\begin{equation}
k_s(t) \simeq
\left(
\frac{\alpha^2 t^2}{L^5(t)}
\right)^{\frac{1}{2-n}} \simeq
L_0^{-1}
\left(
\frac{t}{\tnl }
\right)^{-\frac{1}{2(2-n)}}
\ln^{\frac{5}{4(2-n)}} \left(
\frac{t}{2 \pi \tnl }
\right).
\label{ksasympt}
\end{equation}
Some comments are now made. First, it may seem paradoxical that the
switching wavenumber $k_s(t)$ is not the inverse of the
switching distance $l_s(t)$. The reason is that in physical space we
have a Gaussian competing with a power law, whereas in Fourier space
we have two power laws competing; this renders inapplicable na\"{\i}ve
Fourier phenomenology in which distances are roughly inverse
wavenumbers. Second, let us define an energy wavenumber $k_L(t) =
L^{-1}(t)$, which is roughly the wavenumber around which most of the
kinetic energy resides. From (\ref{Ltasymptotic}) $k_L(t) \sim
(t\sigma_\psi)^{-1/2}$ (ignoring logarithmic corrections). We then
have from (\ref{ksasympt}), still ignoring logarithmic corrections\,:
\begin{equation}
\frac{k_s(t)}{k_L(t)} \sim
\left( \frac{t}{\tnl } \right)^{- \frac{n-1}{2(2-n)}}.
\label{kskLratio}
\end{equation}
Hence, the switching wavenumber goes to zero much faster than the
energy wavenumber\,: in Fourier space the separation of the two regions
is much more manifest. This is why the results of the numerical
simulations in Section~\ref{s:numerics} will be presented in Fourier
space. 

Let us also observe that the ratio of the energy in the outer region
to the total energy, a measure of how well the Kida law is satisfied,
is equal to $(t/\tnl)^{-{3(n-1)\over n-2}}$ (up to logarithms) and thus
becomes very small when $t\gg \tnl$, unless $n$ is very close to unity.

Thus, we have established the central claim of this paper, namely that
for $1<n<2$ there is no globally self-similar evolution of the energy
spectrum. Of course, as $n\to 2$ the inner $k^2$ region overwhelms the
outer $|k|^n$ region and as $n\to 1$ the converse happens, so that in both
instances global self-similarity tends to be reestablished.

\section{Numerical experiments}
\label{s:numerics}

The purpose of this section is to obtain numerical confirmation of the rather unusual
long-time regime predicted by the theory when the spectral exponent $n$ is 
between one and two. There should then be three different power-law regions in the energy 
spectrum\,:
\begin{equation}
\begin{array}{l}
\ds E(k) \propto |k|^{-2}\quad {\rm for}\,\, |k|\gg L^{-1}(t);\\[1.6ex]
\ds E(k) \propto |k|^2\quad {\rm for}\,\, k_s(t)\ll |k|\ll L^{-1}(t);\\[1.6ex]
\ds E(k) \propto |k|^n\quad {\rm for}\,\, |k|\ll k_s(t),
\end{array}
\label{3regions}
\end{equation}
where $k_s(t)$ and $L(t)$ are given by (\ref{ksasympt}) and (\ref{Ltasymptotic}), 
respectively.

Clearly, such simulations require extremely high spatial resolution
since three power-law ranges are present.  Fortunately, for the
Burgers equation there is no need to numerically integrate
(\ref{burgers})\,: advantage can be taken of the maximum
representation (\ref{Phifunction}) to directly construct, from the
initial data, the solution in the limit of vanishing viscosity at any
given time $t$.  The principles of the method have been presented in
detail elsewhere (She, Aurell \& Frisch 1992; Noullez \& Vergassola
1994; Vergassola {\it et al.} 1994; Aurell, Gurbatov \& Simdyankin
1996) and will only be briefly recalled here.

The method exploits two observations. First, the maximum
representation (\ref{Phifunction}) becomes a Legendre transformation
after expansion of the quadratic term. Second, when the problem is
discretized on a grid of $N$ points, the search for the maximizing
Lagrangian point $a(x,t)$ can be done by a dichotomic procedure which
uses the non-decreasing property of the inverse Lagrangian map. The
number of operations can thus be reduced from a brute-force estimate $O(N^2)$
to $O(N\ln_2 N)$.

In the simulations it is of course necessary to introduce both
large-scale and small-scale cutoffs. The small-scale cutoff must be
chosen sufficiently small compared to the integral scale to leave room
for a $k^{-2}$ inertial range. Since the integral scale grows with the
time, this constraint is not very stringent. In practice, we take the
initial integral scale about a factor ten larger than the mesh.

For the large-scale cutoff, we assume periodic boundary conditions
which facilitate Fourier transformations and make no significant
difference as long as we analyse scales much smaller than the spatial
periodicity.

In our simulations, the maximum wavenumber is taken to be $1/2$, so
that wavevectors run from $\pm 1/N$ to $\pm 1/2$, where $N$ is the
resolution, that is the number of grid points in physical space. In
all our simulations we take $N=2^{21}\approx 2\times 10^6$. The energy
spectrum for the initial Gaussian velocity is
\begin{equation}
E_0(k) = \alpha_n^2 |k|^n \mbox{\Large e}^{-\frac{k^2}{2k_0^2}},\quad
k=0, \pm{1\over N}, \pm {2\over N}, \ldots, \pm {1\over 2},
\label{NIS}
\end{equation}
with $k_0 \approx 0.07$ and $\alpha_n$ chosen such that $\langle
\psi_0^2(x)\rangle =1$.  The initial integral scale and nonlinear
time, as defined in (\ref{defL0}) and (\ref{deftnl}), respectively, are then given by
\begin{equation}
L_0 = {1\over k_0}\,\left[{\Gamma\left({n-1\over2}\right)\over \Gamma\left({n+1\over2}\right)}
\right]^{1/2}, \qquad \tnl = {1\over k_0^2}\,
{\Gamma\left({n-1\over2}\right)\over \Gamma\left({n+1\over2}\right)}.
\label{L0tnlnum}
\end{equation}
With the chosen value of $k_0$, the integral scale is around ten and the nonlinear time
around one hundred.

The initial potential is generated by fast Fourier transform (FFT)
from its Fourier coefficients $\hat\psi_0(k)$ which are complex
Gaussian variables of variance $\langle |\hat\psi_0(k)|^2\rangle =
E_0(k)/(Nk^2)$. After the solutions at suitable output times have been
constructed with the procedure described above, they are Fourier
transformed and their energy spectra are calculated by averaging
typically over $10^3$ realizations.\footnote{This Monte Carlo procedure is
easily parallelized in the implementation on an IBM SP2 machine.} 

The precise value of the spectral exponent $1<n<2$ in our runs
depends, of course, on the particular tradeoffs desired\,: the closer
$n$ is to 1, the more conspicuous is the persistence of the initial
$|k|^n$ range; and the closer it is to 2, the more conspicuous is the
(nonlinearly generated) $k^2$ range. A natural unit for the output
times is $t^*=N^2$. Indeed, we know from (\ref{Ltasymptotic}) that the
integral scale grows roughly as $t^{1/2}$. Thus $t^*$ is roughly the
time needed for the integral scale to grow from an initial value
$O(1)$ to a value comparable to the largest scale available in the
simulation, which is $O(N)$. In practice, we take output times
\begin{equation}
t_i = 10^{i-6}\, N^2, \qquad i=1,2, \ldots, 8. 
\label{outputtimes}
\end{equation}

\begin{figure}
\iffigs \centerline{\psfig{file=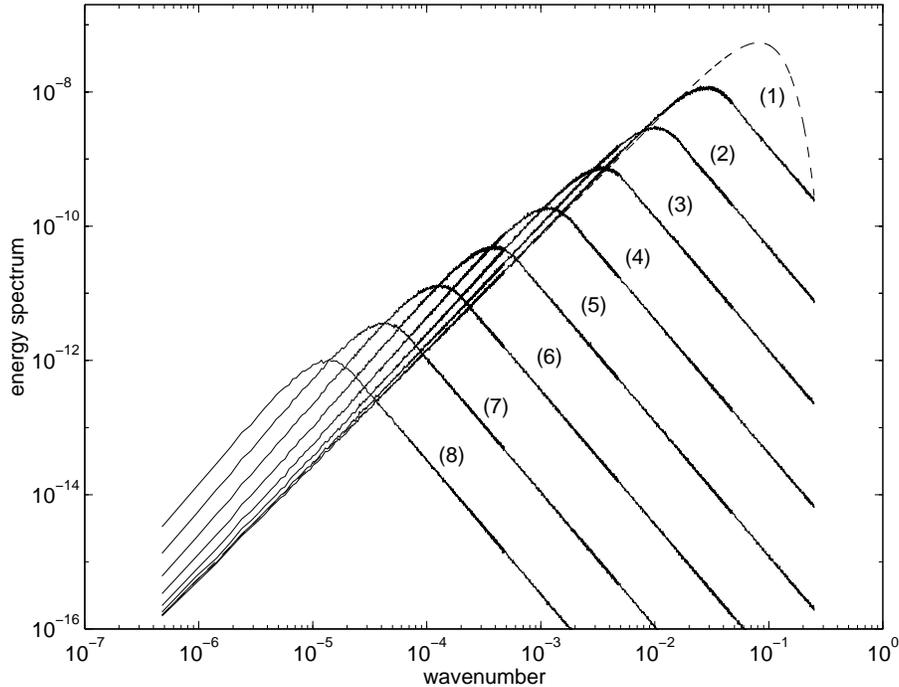,width=12cm,clip=}} \fi
\caption{Evolution of the energy spectrum with an initial spectrum
(dashed line) proportional to $k^n$ ($n=1.7$) at small wavenumbers
$k$. Resolution $N=2^{21}$. Spectra averaged over $10^3$
realizations. The labels correspond to output times $t_i=10^{i-6}\,N^2$,
($i=1,2,\ldots 8$).}
\label{f:spectral-curves}
\end{figure}

We describe now the results of our numerical experiments.
Fig.~\ref{f:spectral-curves} shows the evolution of the spectrum for
$n=1.7$ and is mostly meant to bring out qualitative aspects. At large
wavenumbers a $k^{-2}$ inertial range develops very
quickly.\footnote{The time scale for shocks to develop is $\tnl$.} At
short times the spectrum preserves its initial value at all but the
largest wavenumbers, in accordance with the permanence of large
eddies. The wavenumber where the spectrum achieves its maximum, which
we use as an operational measure of the inverse integral scale at time
$t$,\footnote{It is actually almost exactly equal to this (see
Gurbatov, Malakhov \& Saichev 1991, Fig.~5.13).} moves to smaller and
smaller wavenumbers. At large times the spectrum appears to change in
a self-similar way.

\begin{figure}
\iffigs
\centerline{\psfig{file=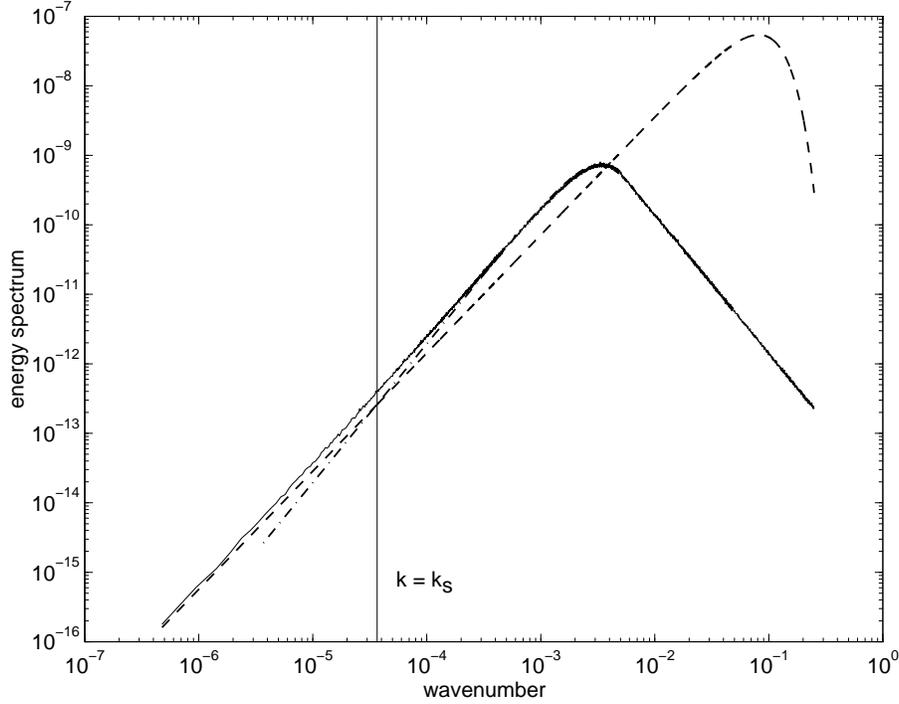,width=12cm,clip=}} \fi
\caption{Energy spectrum at $t=t_3=10^{-3}\,N^2$. Same conditions as
Fig.~\protect\ref{f:spectral-curves}. Initial spectrum\,: dashed
line. Leading-order asymptotic theory in inner region\,: dashed dotted
line. The vertical line shows the switching wavenumber.}
\label{f:one-spectral-curve}
\end{figure}

We turn to a more quantitative description.
Fig.~\ref{f:one-spectral-curve} shows the energy spectrum at the third
output time $t_3=10^{-3} \, N^2$. We also plot the initial spectrum
with its $|k|^n$ tail which should be preserved at small wavenumbers
(leading term of outer expansion) and the leading term of the
inner  expansion, obtained by Fourier transformation of
(\ref{Bvasymptotic}), which has a $k^2$ tail at small wavenumbers.
The intersection of the latter two is the switching wavenumber
$k_s(t)$ as defined by (\ref{ksasympt}).
It is seen that, except in the immediate neighborhood of the switching wavenumber, the leading
terms of the inner and outer expansions are in very good agreement with our simulations.

\begin{figure}
\iffigs
\centerline{\psfig{file=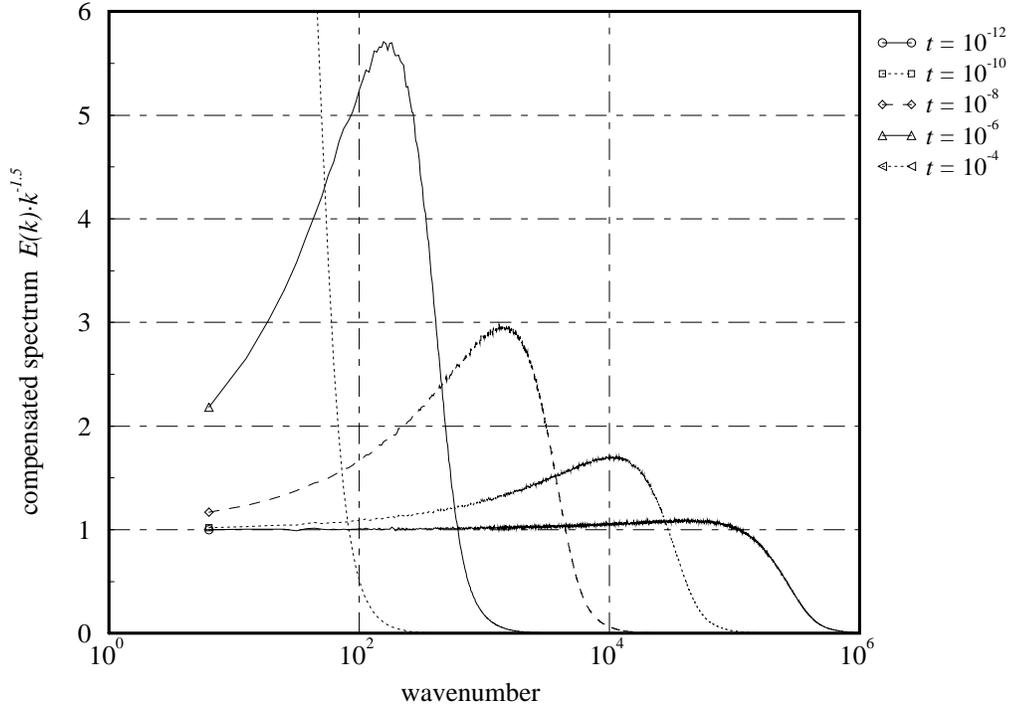,width=10cm}}
\vspace{12mm} \fi \caption{Evolution of the energy spectrum compensated by a
$k^{-n}$ factor in order to reveal the region of ``permanence of large
eddies'' (PLE) as a plateau with unit height. Spectral exponent
$n=1.5$. Resolution $N=2^{20}$. Spectra averaged over 30,000
realizations. The unit of time is different from the other figures,
the spatial period being here unity. Observe  the progressive shrinking of the
PLE zone. ({\em Courtesy A.~Noullez.})}
\label{f:comp1.5} \end{figure}

Fig.~\ref{f:comp1.5}, communicated by A.~Noullez, gives additional evidence for the permanence
of large eddies.

\begin{figure}
\iffigs
\centerline{\psfig{file=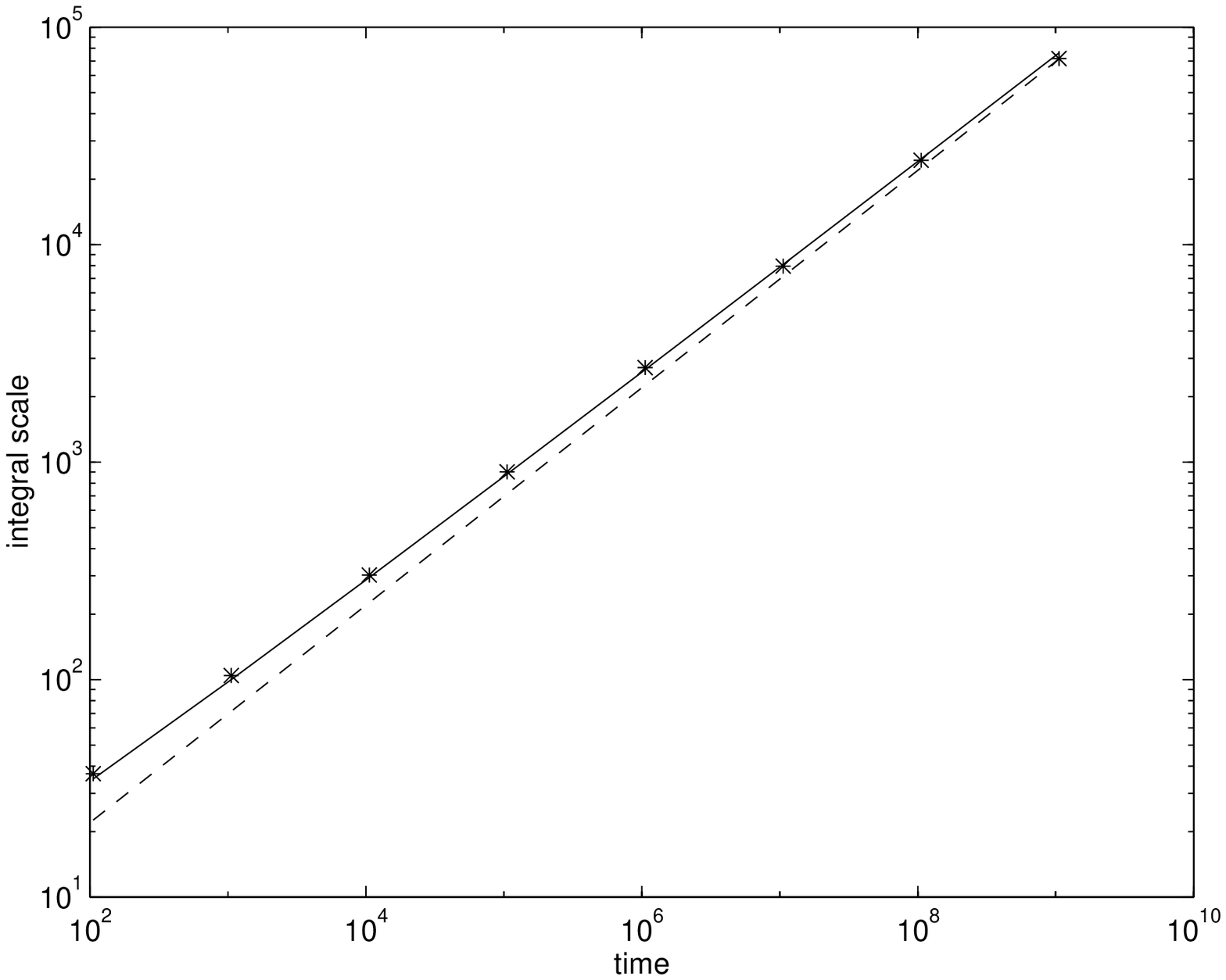,width=10cm,clip=}} \fi \caption{Evolution
of the computed integral scale $L(t)$ (asterisks) compared to the
theoretical leading-order prediction (solid line) and the same without
the logarithmic correction (dashed line). Same conditions as
Fig.~\protect\ref{f:spectral-curves}.}
\label{f:lt-curve}
\end{figure}

\begin{figure}
\iffigs
\centerline{\psfig{file=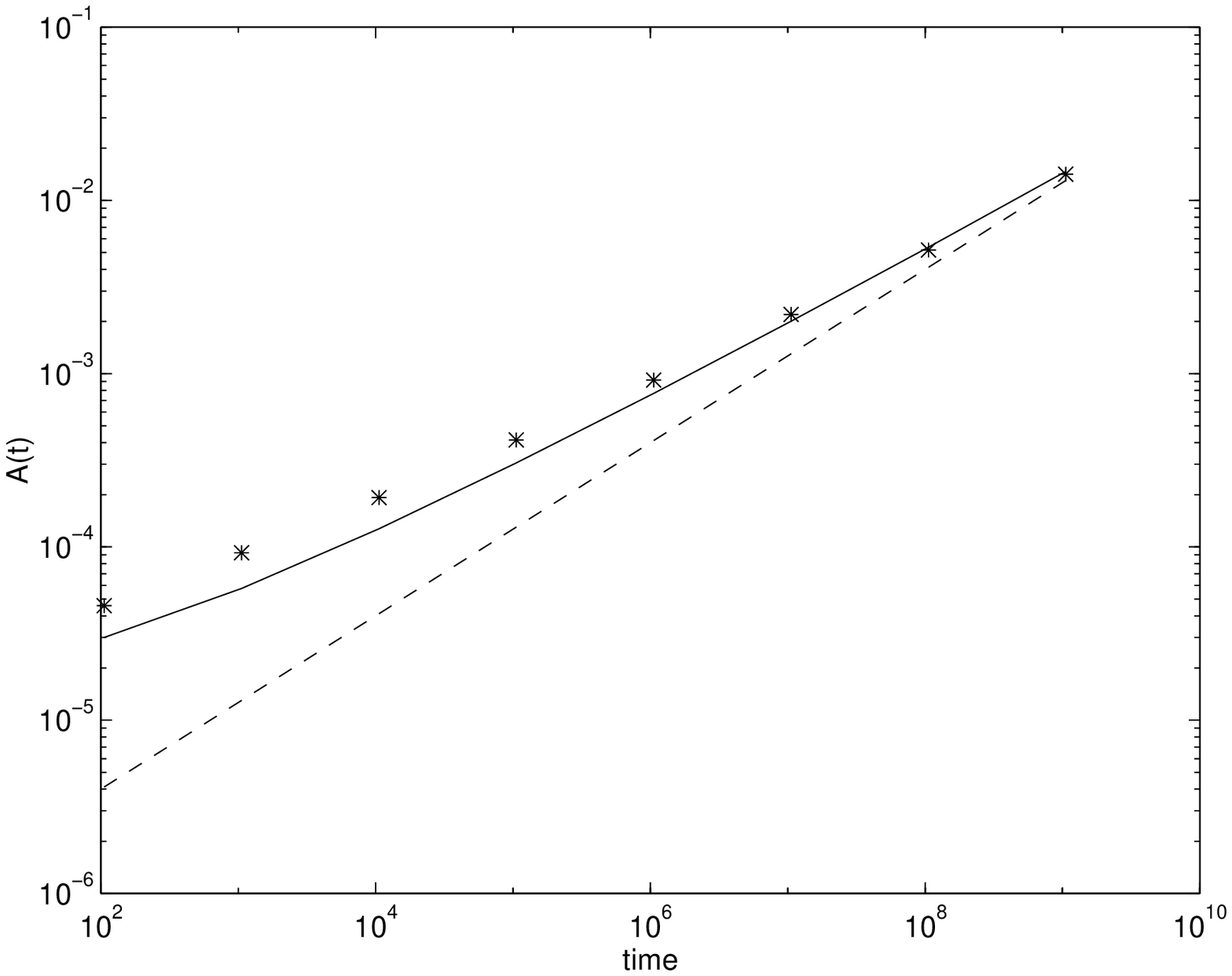,width=10cm,clip=}} \fi\caption {Evolution of the coefficient $A(t)$ in front of the $k^2$ term
at intermediate wavenumbers (asterisks), compared to the leading-order
theoretical prediction (\protect\ref{Aoft}) (solid line) and the same
without its logarithmic correction (dashed line).  Same conditions as
Fig.~\protect\ref{f:spectral-curves}.}
\label{f:At-curve}
\end{figure}

We examine now the temporal evolution of various quantities for which
we have theoretical predictions.  Fig.~\ref{f:lt-curve} shows the
evolution of the integral scale compared to the theoretical
prediction(\ref{Ltasymptotic}). The agreement is excellent provided
that the logarithmic correction is taken into account. By (\ref{uvrms}) and
(\ref{uLt}) this also implies the validity of Kida's log-corrected law for
the energy decay.
Fig.~\ref{f:At-curve} shows the corresponding information for the
coefficient
$A(t)$.

\begin{figure}
\iffigs
\centerline{\psfig{file=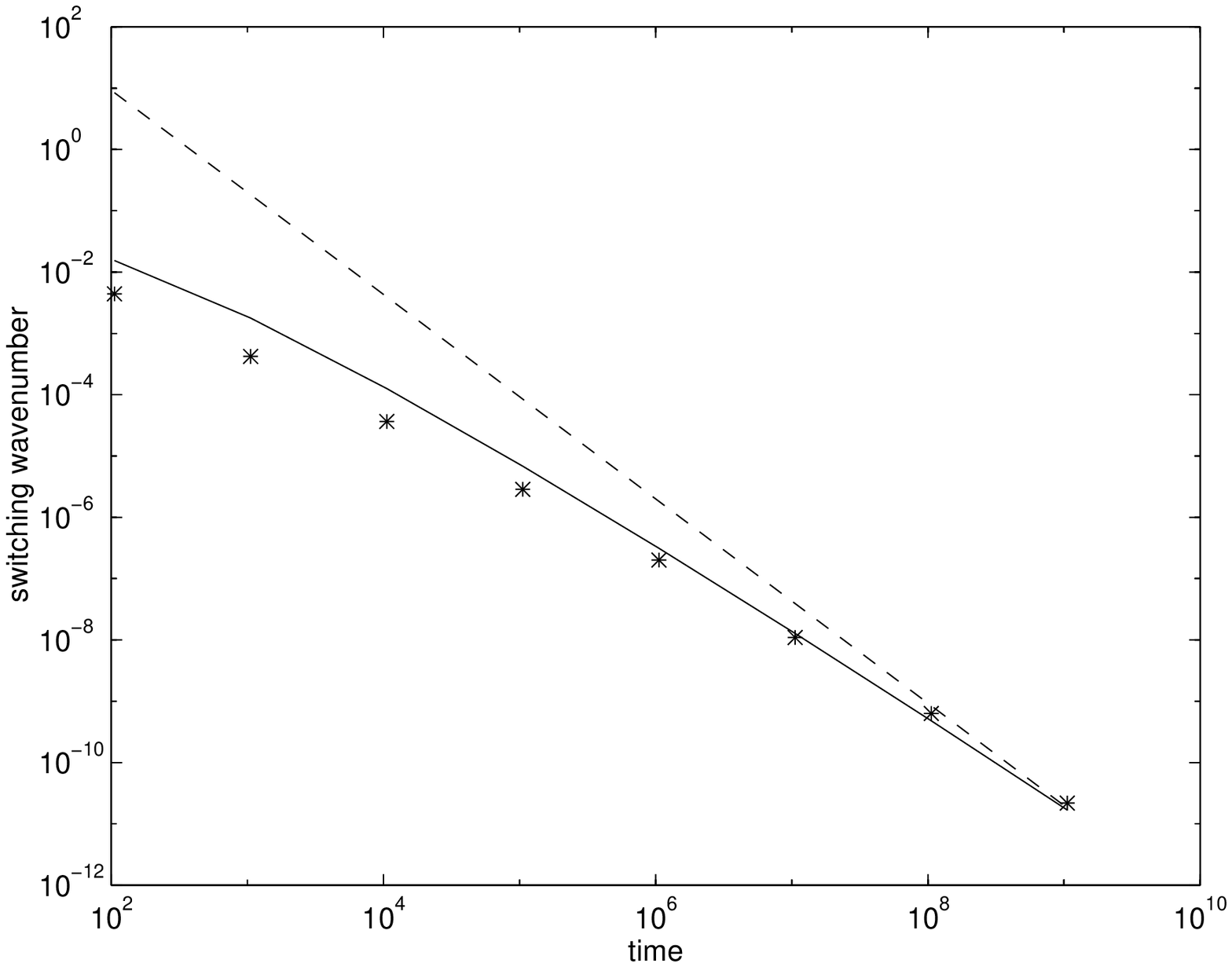,width=10cm,clip=}} \fi \caption{Evolution
of the switching wavenumber $k_s(t)$, compared to the leading-order
theoretical prediction (\protect\ref{ksasympt}) (solid line) and the
same without its logarithmic correction (dashed line).  Same
conditions as Fig.~\protect\ref{f:spectral-curves}.}
\label{f:kc-curve}
\end{figure}

\begin{figure}
\iffigs
\centerline{\psfig{file=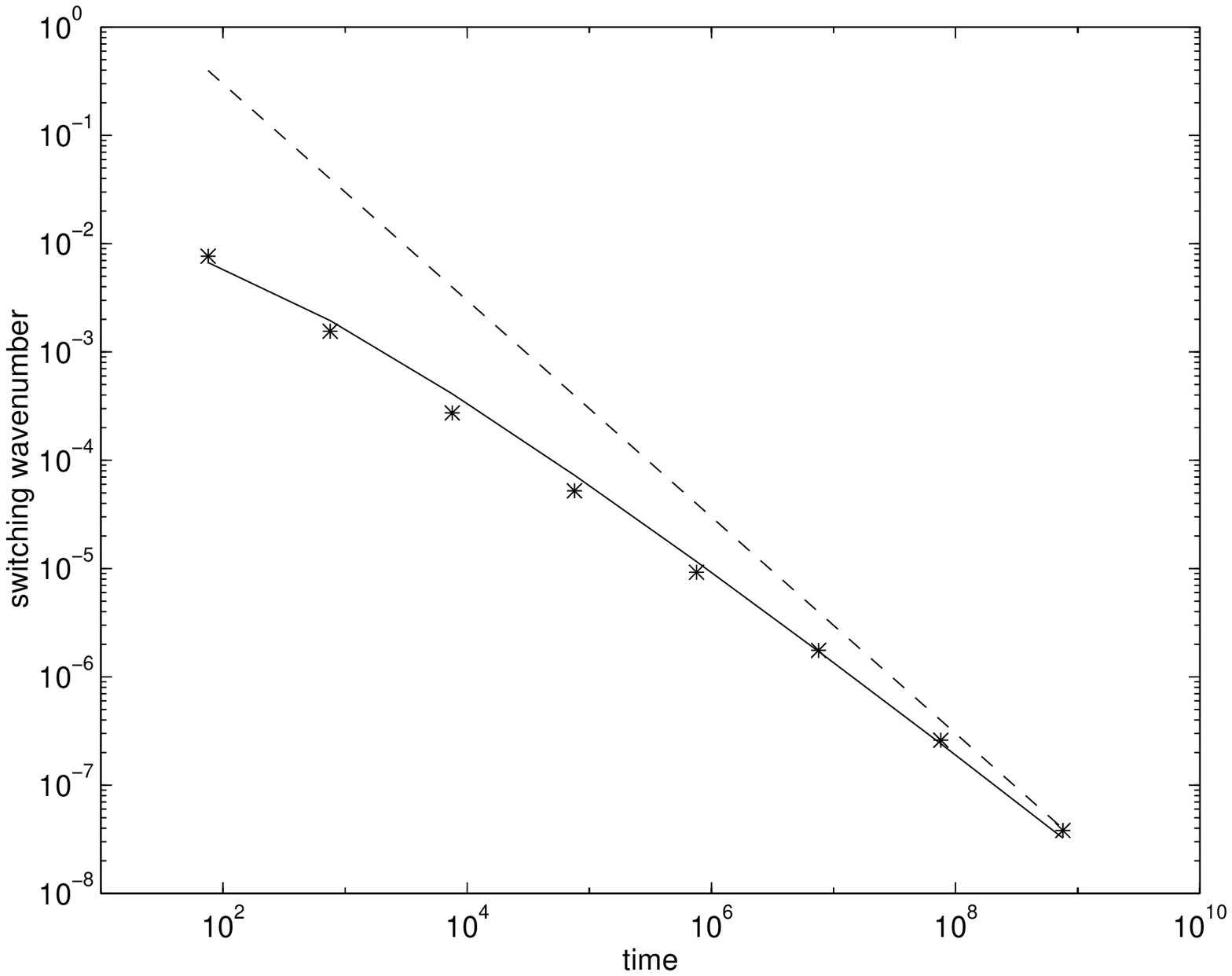,width=10cm,clip=}}
\fi \caption{Evolution
of the switching wavenumber $k_s(t)$ for $n=1.5$. Otherwise, as in 
Fig.~\protect\ref{f:kc-curve}.}
\label{f:kc-curve-1.5}
\end{figure}

Fig.~\ref{f:kc-curve} shows the evolution of the switching wavenumber
$k_s(t)$, compared  with the theory (\ref{ksasympt}).
Finally, Fig.~\ref{f:kc-curve-1.5} shows the evolution of the
switching wavenumber, but now for a spectral index $n=1.5$. The
agreement with the leading-order theory is slightly better than
for $n=1.7$.

\section{Concluding remarks}
\label{s:conclusion}

The main conclusions have already been presented in
Section~\ref{s:intermediate} (see also the Abstract). Here, we make
additional comments. 

We have given an explicit formula only for the leading-order term of
the law of energy decay. The higher-order terms have just been
bounded\,: they appear to be smaller (in relative terms) by negative
powers of the time $t$. It would be useful to have explicit
expressions for the subdominant terms. 

Our derivation of the law of energy decay uses an
asymptotic expression for the probability of non-crossing of the initial
potential and suitable parabolas. Kida's (1979) original derivation
obtained this from the distribution of shocks, which he related to the
distribution of (high local)  maxima in the initial potential. Actually, it
should be possible to reformulate our theory in terms of maxima rather
than of upcrossings. Roughly, we expect that high-level upcrossings
should occur  very close to high-level maxima. This statement cannot
be exactly true since two successive up- and downcrossings of the
(initial) potential with  a parabola need not be separated by a maximum
of the potential. However, crossings of a parabola having  a high-lying
apex are very likely to happen in the flat region near the apex
of the parabola where up- and downcrossings are separated by a
maximum. Quantifying such remarks would be of special interest 
for extension of the theory to the multidimensional Burgers
equation. Indeed,  in $d>1$ space dimensions, the equivalent of the
crossings are $(d-1)$-dimensional manifolds for which the extension of
the method presented here is far from straightforward. But (local)
maxima are still point processes and can be handled via the
$d$-dimensional version of the Balian--Schaeffer formula (Balian \&
Schaeffer 1989). 

There are a number of other interesting extensions of the
present work, such as the effect of viscosity and of non-Gaussian
initial potentials. Instances of these have already been studied 
by Funaki, Surgailis \& Woyczynski (1995), Surgailis (1996), Hu \&
Woyczynski (1996) and Surgailis (1997).

Finally, a completely open question is the extension of the present
ideas to three-dimensional {\it incompressible\/} turbulence. Does the
mechanism which prevents globally self-similar decay for $1<n<2$ have
a counterpart for the Navier--Stokes problem\,? 

\vspace{2mm} \par\noindent {\bf Acknowledgements.}  We have benefited
from discussions with A.~Noullez. W.A.~Woyczynski and S.~Lehr have
kindly sent us various papers indicated by a referee. This work was
supported by RFBR-INTAS through grant 95-IN-RU-0723 (S.N.G., S.I.S.,
U.F. and E.A.), by RFBR through grant 96-02-19303 (S.N.G. and S.I.S.),
by the Swedish Institute (S.I.S.), by the Swedish Natural Science
Research Council through grant S-FO-1778-302 (E.A.), by a grant from
the {\em Fondation des Treilles\/} (E.A., U.F., S.N.G. and G.T.), by
the French Ministry of Higher Education (S.N.G. and G.T.) and by the
Hungarian Science Foundation grant OTKA F-4491 (G.T.). S.~Gurbatov and
G.~T\'oth thank the Observatoire de la C\^ote d'Azur, S.~Simdyankin
and E.~Aurell the Center for Parallel Computers, and E.~Aurell the
Radiophysical Department of Nizhny Novgorod University, for
hospitality and creating nice and pleasant working conditions.

\newpage
\section*{Appendix} 
\appendix
\renewcommand{\theequation}{\thesection.\arabic{equation}}
\section{The Balian--Schaeffer formula for ``particle'' distributions}
\setcounter{equation}{0}
\label{a:bs}

The original work of Balian \& Schaeffer (1989) was concerned with the
following type of problem. One has a distribution of ``particles''
(galaxies, in the applications considered by them) in the
three-dimensional space which is homogeneous, i.e.\
translation-invariant and one wants to determine the probability that
there is no particle at all within some {\em finite\/} domain. For the
application to the Burgers equation, the role of the particles will be
played by the upcrossings between the initial potential and a given
parabola, which constitute a point process in the sense of Leadbetter,
Lindgren \& Rootzen (1983). Although the potential is a homogeneous
random function, the presence of the parabola makes our problem
inhomogeneous. Furthermore, our domain is the whole real line and is
thus not bounded.\footnote{The formalism developed hereafter applies
identically to bounded and unbounded domains. The integral sign will
always denote a simple or multiple integration over the whole domain.}
Nevertheless, our approach will remain rather close in spirit to the
original Balian--Schaeffer approach. In the remainder of this appendix
we shall use the word ``particle'' rather than ``intersection'', since
the formalism has obviously applications beyond the Burgers equation.

Let there be given a random set of particles on the real line. The
total number of particles is assumed to be random but almost surely
finite. We are actually interested in finding the probability of having
no particle at all. We shall characterize this random set by the
so-called $n$-particle densities. $F_1(y_1)\,dy_1$ is the probability
of having a particle within the interval $[y_1,y_1+dy_1]$. Similarly,
$F_n(y_1,y_2,\ldots,y_n)\,dy_1dy_2\ldots dy_n$ is the probability of
having (at least) $n$ different particles within, respectively, the
intervals $[y_1,y_1+dy_1]$, $[y_2,y_2+dy_2]$, \ldots and, finally,
$[y_n,y_n+dy_n]$. Note that when the particles are independent (an
assumption not made here), $F_n$ is just a product of $F_1$ factors.
Note also that the $F_n$\ms1 are not normalized. For example $\int
F_1(y_1)dy_1$ is the mean number of particles.\footnote{This follows
from the observation that the mean number of particles in the interval
$[y_1,y_1+dy_1]$ differs by $O(dy_1^2)$ from the probability of
having {\em one\/} particle within the interval $[y_1,y_1+dy_1]$.}
Balian and Schaeffer's work leads to expressions for $p_n$, the
probability of having exactly $n$ particles in the domain, in terms of
the $F_n$\ms1. We shall here concentrate on the evaluation of $p_0$,
the probability of having no particle at all.

We begin by introducing the exclusive $n$-particle densities
corresponding to the case where there are exactly $n$ particles.
$P_n(y_1,y_2,\ldots,y_n)$ $dy_1dy_2\ldots dy_n$ is the probability of
having a particle within the interval $[y_1,y_1+dy_1]$, another particle
within the interval $[y_2,y_2+dy_2]$, \ldots, a particle within the
interval $[y_n,y_n+dy_n]$ and {\em no other particle}. Observe that
\begin{equation}
\int  P_n(y_1,y_2,\ldots,y_n)\,dy_1dy_2\ldots dy_n= n! \,p_n.
\label{normPn}
\end{equation}
The factor $n!$ is present because the $P_n$\ms1 are here defined for
{\em unlabeled\/} particles.
We shall use (\ref{normPn}) to calculate $p_n$ for $n=1,2,..$ and then obtain
$p_0$ from the obvious relation
\begin{equation}
p_0= 1- p_1 - p_2 -\ldots p_n -\ldots,
\label{pzerofrompn}
\end{equation}
expressing that the total probability is unity.

We show now that the $p_n$\ms1 may be expressed in terms of the
$F_n$\ms1. First we relate the $F_n$\ms1 to the $P_n$\ms1. To have a
particle in $[y_1,y_1+dy_1]$ we must have one of the following
mutually exclusive situations\,: (i) a total of exactly one particle
in $[y_1,y_1+dy_1]$, (ii) a total of exactly two particles, one of
them in $[y_1,y_1+dy_1]$ and the other one at any place, \ldots.
Hence,
\begin{eqnarray}
F_1(y_1) &=& P_1(y_1) +\int P_2(y_1,y_2)\, dy_2  \nonumber \\
& &  +{1\over{2!}}\int
P_3(y_1,y_2,y_3)\,dy_2dy_3 + \ldots.
\label{F1byPn}
\end{eqnarray}
The presence of the factorials in the denominators comes again from the
unlabeled character of the particles.
Similarly, we have
\begin{equation}
F_n(y_1,\ldots y_n) = \sum_{l=0}^{\infty} {1\over {l!}}
\int P_{n+l}(y_1,\ldots y_n, y_{n+1},
\ldots,y_{n+l}) dy_{n+1}\ldots dy_{n+l}.
\label{FnbyPn}
\end{equation}
Next we introduce the quantities
\begin{equation}
f_n = \int F_n(y_1,y_2,\ldots,y_n)dy_1dy_2\ldots dy_n .
\label{fn}
\end{equation}
Integrating (\ref{F1byPn}) over $dy_1$, we obtain
\begin{equation}
f_1=p_1+2!p_2+{3!\over 2!} p_3 +\ldots {m!\over(m-1)!}p_m +\ldots.
\label{f1pn}
\end{equation}
More generally, integrating (\ref{FnbyPn}) over $dy_1$,
\ldots, $dy_n$, we have
\begin{equation}
f_n=n!p_n+(n+1)!p_{n+1}+\ldots {m!\over(m-n)!}p_m +\ldots.
\label{fnpn}
\end{equation}

Multiplying (\ref{fnpn}) by $(-)^n/n!$ and summing over $n$, we obtain
\begin{equation}
p_0= \sum_{n=0}^{\infty} {{(-)^n}\over{n!}}f_n.
\label{pzerofn}
\end{equation}
Let us briefly digress and consider the case of independent particles,
when $F_n(y_1,\ldots,y_n) = F_1(y_1)F_1(y_2)\cdots F_1(y_n)$, and
thus $f_n=f_1^n$. From (\ref{pzerofn}) we then obtain
\begin{equation}
p_0= \exp\left(-f_1\right) .
\label{pzeropoisson}
\end{equation}
Since $f_1=\int F_1(y_1) dy_1$ is the mean number of particles,
(\ref{pzeropoisson}) may be viewed as just a consequence of the Poisson
distribution. It is referred to in this paper as the ``Poisson
approximation''.

The general case, of correlated particles, is handled by introducing
correlation functions through a cluster expansion\,:
\begin{eqnarray}
F_1(y_1) &=& C_1(y_1) \label{FC1}\\
F_2(y_1,y_2) &=& C_1(y_1)C_1(y_2)+ C_2(y_1,y_2) \label{FC2}\\
&\vdots& \nonumber \\
F_n(y_1,y_2,\ldots) &=& \sum_{\hbox{all partitions}}
\cdots C_{l}(y_{r_1},\ldots,y_{r_l})\cdots
\label{cluster}
\end{eqnarray}
We also need the cumulants
\begin{equation}
c_n =\int C_n(y_1,y_2,\ldots,y_n)dy_1dy_2\ldots dy_n.
\label{cn}
\end{equation}
Note that the $c_n$\ms1 for $n>1$ vanish in the case of independent
particles.  The name ``cumulant'' for the $c_n$\ms1 is justified by
the observation that the $c_n$\ms1 and the $f_n$\ms1 defined in
(\ref{fn}) are related like moments and cumulants\,: if we introduce
the generating function
\begin{equation}
\Omega(z) = \sum_{n=0}^\infty {z^n\over n!}f_n,
\label{defOmega}
\end{equation}
we easily obtain from (\ref{FC1})-(\ref{cluster})
\begin{equation}
\Omega(z) = \exp\left(\sum_{m=1}^\infty {z^m\over m!}c_m\right).
\label{Omegaexp}
\end{equation}
Expansion of the r.h.s.\ of (\ref{Omegaexp}) in powers of $z$ and
identification with the expansion (\ref{defOmega}) allows the
successive determination of the set of $f_n$\ms1 in terms of the set
of $c_n$\ms1 and, conversely, identification of the expansion of the
logarithm of the r.h.s.\ of (\ref{defOmega}) with the logarithm of
(\ref{Omegaexp}) gives the $c_n$\ms1 in terms of the $f_n$\ms1. In
particular, we have
\begin{eqnarray}
c_1&=&f_1 = \int F_1(y_1)\,dy_1, \label{f1c1}\\
c_2&=&f_2 -f_1^2 =
\int \left(F_2(y_1,y_2)-F_1(y_1)F_1(y_2)\right)\,dy_1dy_2 .
\label{c1c2f1f2}
\end{eqnarray}

Comparing  (\ref{pzerofn}) and (\ref{defOmega}), we have $p_0 =
\Omega (-1)$. Then, using  (\ref{Omegaexp}), we obtain the required
expression for the probability of having zero particles\,:
\begin{equation}
p_0 = \exp\left(\sum_{n=1}^\infty {(-)^n\over n!}c_n\right).
\label{pzerocn}
\end{equation}
Of course, if we truncate the series to its first term, we recover
the Poisson approximation.

\newpage
\vspace*{-3cm}
\par\noindent FIGURE CAPTIONS
\vspace{2mm}

\begin{itemize}

\item[Figure 1:] Sketch of the energy spectrum at long times when
$1<n<2$.

\item[Figure 2:] Intersections of the initial potential $\psi_0(y)$
(curve $G_1$)  with the parabola $G_2$.

\item[Figure 3:] Intersections of the initial potential $\psi_0$ with
two parabolas.

\item[Figure 4:] The different regions of integration appearing in
(\protect\ref{c2intexpr}). Near the diagonal $y_1=y_2$ we have regions I and I'
of (\protect\ref{IandII}). Away from the diagonal we have regions II
and II'.

\item[Figure 5:] Evolution of the energy spectrum with an initial spectrum
(dashed line) proportional to $k^n$ ($n=1.7$) at small wavenumbers
$k$. Resolution $N=2^{21}$. Spectra averaged over $10^3$
realizations. The labels correspond to output times $t_i=10^{i-6}\,N^2$,
($i=1,2,\ldots 8$).

\item[Figure 6:] Energy spectrum at $t=t_3=10^{-3}\,N^2$. Same
conditions as Fig.~\protect\ref{f:spectral-curves}. Initial
spectrum\,: dashed line. Leading-order asymptotic theory in inner
region\,: dotted line. The vertical line shows the switching wavenumber.

\item[Figure 7:] Evolution of the energy spectrum compensated by a
$k^{-n}$ factor in order to reveal the region of ``permanence of large
eddies'' (PLE) as a plateau with unit height. Spectral exponent
$n=1.5$. Resolution $N=2^{20}$. Spectra averaged over 30,000
realizations. The unit of time is different from the other figures,
the spatial period being here unity. Observe  the progressive shrinking of the
PLE zone. ({\em Courtesy A.~Noullez.})

\item[Figure 8:] Evolution
of the computed integral scale $L(t)$ (asterisks) compared to the
theoretical leading-order prediction (solid line) and the same without
the logarithmic correction (dashed line). Same conditions as
Fig.~\protect\ref{f:spectral-curves}.

\item[Figure 9:] Evolution of the coefficient $A(t)$ in front of the
 $k^2$ term at intermediate wavenumbers (asterisks), compared to the
 leading-order theoretical prediction (\protect\ref{Aoft}) (solid
 line) and the same without its logarithmic correction (dashed line).
 Same conditions as Fig.~\protect\ref{f:spectral-curves}.

\item[Figure 10:] Evolution of the switching wavenumber $k_s(t)$,
compared to the leading-order theoretical prediction
(\protect\ref{ksasympt}) (solid line) and the same without its
logarithmic correction (dashed line).  Same conditions as
Fig.~\protect\ref{f:spectral-curves}.

\item[Figure 11:] Evolution
of the switching wavenumber $k_s(t)$ for $n=1.5$. Otherwise, everything as in 
Fig.~\protect\ref{f:kc-curve}.

\end{itemize}

\end{document}